\documentclass[twocolumn, trackchanges]{aastex701}
\makeatletter
\long\def\frontmatter@title@above{}
\makeatother

\usepackage{comment}
\usepackage{csquotes}
\usepackage{amsmath}
\usepackage{graphicx}
\graphicspath{{figs/}}
\usepackage{booktabs}
\usepackage{algorithm}
\usepackage{algpseudocode}

\usepackage[dvipsnames]{xcolor}
\usepackage{listings}

\lstdefinestyle{pythonstyle}{
    language=Python,
    basicstyle=\ttfamily\small,
    keywordstyle=\color{purple}\bfseries,
    commentstyle=\color{gray},
    stringstyle=\color{teal},
    numbersep=8pt,
    frame=single,
    breaklines=true,
    showstringspaces=false,
    tabsize=4,
    captionpos=b, 
    frame = none}

\newcommand{\torb}{T_{\rm orb}}
\newcommand{\tisco}{T_{\rm ISCO}}
\newcommand{\tcoal}{t_{\rm coal}}
\newcommand{\dphi}{|\delta\varphi|}
\newcommand{\rr}{\varepsilon}
\newcommand{\Msun}{M_\odot}
\newcommand{\qp}{q_+}
\newcommand{\qm}{q_-}

\newcommand{\chione}{\chi^{(1)}}
\newcommand{\chitwo}{\chi^{(2+)}}

\shorttitle{\texttt{[SIMFONIA $||$ SINFONIA]} - Symplectic, Slimplectic and Magnusian (neural) flows}
\shortauthors{Gomes Da Silva, Lidia Joana}

\begin{document}
\title{Introducing \texttt{SINFONIA}: \underline{Sy}mplectic, slimplectic and Magnusian (\underline{N}eural) \underline{F}lows for \underline{O}rbital \underline{N}umerical \underline{I}ntegration and \underline{A}cceleration \footnote{\texttt{SI/SIM} and not \texttt{SY/SYM} as we keep the pronunciation of the word symplectic in Portuguese, i.e. simpl\'ectico.}}

\author[orcid=0000-0001-9593-4217,sname='Lidia Joana']{Lidia J. Gomes Da Silva}
\affiliation{SISSA, Via Bonomea 265, 34136 Trieste, Italy and INFN Sezione di Trieste}
\affiliation{IFPU - Institute for Fundamental Physics of the Universe,
Via Beirut 2, 34014 Trieste, Italy}
\email[show]{lidiajoana@pm.me $\vert\vert$ Cyberspace: \href{https://lidiajgomesdasilva.io}{lidiajgomesdasilva.io}}

\begin{abstract}
Long-duration gravitational-wave modelling must resolve fast orbital motion together with slow dissipative evolution while preventing small numerical errors from accumulating into secular phase drift. Here we ask whether the finite-time evolution map itself can be learned as an explicit, differentiable, structure-preserving object and then repeatedly composed through a complete inspiral. We construct three neural-flow architectures: a symplectic flow on Galley's doubled phase space, \texttt{[SINFONIA-J0]}; a Taylor-anchored flow, \texttt{[SINFONIA-J1]}; and a Magnusian flow that learns the finite-time dissipative correction in the interaction picture, \texttt{[SINFONIA-J2]}. Applied to a 2.5PN neutron-star inspiral, all three expose the same controlling mechanism: long-time accuracy is governed not by pointwise map error alone, but by its signed projection onto a single secular channel fixed by energy--angular-momentum balance. Encoding this structure allows the learned maps to remain accurate through $10^{2}$--$10^{5}$ window compositions to coalescence at timesteps of a full orbital period and beyond, reaching chained phase errors orders of magnitude below a benchmark slimplectic integrator at lower cost. The same secular structure can also be exploited for physics inference: when the channel is left unconstrained, the accumulated phase retains enough information to recover an un-modelled dynamical-friction-like force, both parametrically and as a learned function of separation. Network-off controls isolate the contribution of learning from the analytic structure already built into each map. These results establish a proof of concept for structure-preserving learned evolution maps as tools for fast long-duration integration and physics inference in gravitational-wave source modelling.
\end{abstract}


\section{Introduction}\label{sec:intro}
Long-term integration is central to the modelling of astrophysical binaries. Compact binaries may execute many thousands of orbital cycles before merger, while radiation reaction drives a slow secular evolution of the orbital energy, angular momentum, and gravitational-wave phase. Accurate waveform modelling \citet{lisa2025waveform, colpi2024lisa, abac2026science, gupta2025possible} therefore requires numerical schemes that resolve dynamics across widely separated timescales while distinguishing the physical secular evolution from spurious numerical drift. Adaptive high-order integrators are routinely employed to control local truncation error as the orbital timescale evolves; nevertheless, local error control alone does not guarantee faithful long-time evolution of the underlying geometric structure or accumulated orbital phase.  This issue becomes particularly important in applications requiring repeated integrations across large regions of parameter space, where the cost of conventional time stepping must be paid anew for each system. Geometric numerical integration instead incorporates properties of the dynamics directly into the evolution map. For conservative Hamiltonian systems, symplectic integrators preserve the symplectic structure and consequently exhibit favourable long-time behaviour \cite{hairer2013geometric}. A symplectic integrator is not, in general, more accurate than a Runge--Kutta method of the same order over a single fixed-step; its advantage emerges under repeated composition. Over very long integrations, preservation of the symplectic form typically prevents the secular energy drift exhibited by generic integrators, replacing it with a bounded, oscillatory error associated with a nearby modified Hamiltonian. What the method \emph{preserves}, rather than what it resolves on any one step, is what survives composition. For a particularly clear demonstration of this behaviour in post-Newtonian spinning binary dynamics, we refer the reader to Figs.~1--5 of \citet{lubich2010symplectic}.\footnote{We have independently reproduced the \texttt{C} \texttt{SPN} integrator of \citet{lubich2010symplectic} in \texttt{JAX}, obtaining comparable accuracy and computational performance. Additionally we extended it to higher-orders; details and validation tests are provided in \citep{Simfonia_Sinfonia_PN_EOB}. Preliminary results available in Slides 12-17 of \citep{GomesdaSilva2026sinfonia}.}

This picture changes once radiation reaction is included. Dissipative dynamics is no longer Hamiltonian \emph{on the physical phase space}, and ordinary symplecticity is therefore not the appropriate structure to preserve. One solution is to admit a splitting framework and compose symplectic integrators with symmetric integration methods as demonstrated by \citet{lubich2010symplectic, heinze2026n}. While no longer symplectic the integrators retain key long term advantages of the symplectic integrator they are composed with. Alternatively, slimplectic integrators extend the variational construction to such systems through the doubled-variable non-conservative action principle, allowing quantities such as energy and angular momentum to evolve consistently with the physical dissipation \cite{galley2013classical,galley2014principle,tsang2015slimplectic, slimplecticcode}.
Complementary generator-based descriptions encode finite-time evolution directly. The Magnus expansion \citep{magnus1954exponential} solves a linear operator equation by exponentiating a generator built from nested commutators. Its classical counterpart replaces these by nested Poisson brackets, so that the generator acts on phase space and its exponential is a canonical map: \citet{kim2025classical} construct it for classical scattering as the generator of the canonical transformation from in- to out-states, and \citet{kim2026magnusian} name it the \emph{Magnusian} and distinguish it from the eikonal phase and the on-shell action. Both are formulated for scattering. \citet{blanco2026magnusian} carries the construction to bound, finite-time evolution and to dissipative dynamics, and applies it to leading-order gravitational radiation reaction \emph{for the first-time}. 

A slimplectic integrator nevertheless remains a local numerical object: given the state at $t_n$, it solves the \emph{implicit} discrete variational equations for the state at $t_{n+1}$, one step at a time and along one trajectory. A complementary description of the same dynamics is the finite-time \emph{flow map}
\begin{equation}
  \Phi_h : z(t) \longmapsto z(t+h),
  \label{eq:flowmap}
\end{equation}
defined on a region of phase space and thereafter advanced by composition. Finite-time maps are, of course, classical objects: Lie-series methods, Wisdom--Holman mappings \citep{wisdom1991symplectic}, and variational integrators \citep{marsden2001discrete} all construct dynamics at the level of maps. The machine-learning (ML) question is different: \emph{can such a map be represented by a trainable, explicit, differentiable function while retaining the structural properties responsible for good long-time behaviour?}

Learning the flow map also changes the computational cost. An expensive training stage is performed once, after which the learned map can be evaluated repeatedly for many initial conditions and/or parameter choices, amortizing the cost of numerical time-integration. Neural operators and related learned dynamical models exploit precisely this idea \citep{lu2019deeponet,cardoso2025exactly,li2020fourier,duruisseaux2025fourier}. The difficulty is that a generic neural approximation of a vector field or solution operator need not inherit the geometry of the underlying dynamics, and small structural errors can accumulate catastrophically under repeated composition. We refer the reader to \citet{GomesdaSilva2026sinfonia} where several \underline{g}ravity-\underline{in}formed \underline{n}eural \underline{n}etworks (\texttt{GINNNs}) are tested on a simpler Schwzarschild Binet toy-model. \footnote{Specifically, the implementation of this toy in different \texttt{GINNNs} architectures can be found in Slides 21-29.} Structure-preserving learning addresses this problem by placing part of the physics in the parametrization itself. Hamiltonian neural networks
\citep{greydanus2019hamiltonian} and Lagrangian neural networks \citep{cranmer2020lagrangian} impose structure at the level of the governing generator, while \texttt{SympNets} \citep{jin2020sympnets}, symplectic neural flows \citep{canizares2024symplectic,xu2026cosynflow}, and Taylor-anchored flows \citep{fang2025learning,fang2025accelerating} construct or constrain the finite-time map itself. In these approaches the structural constraint is built into the representation, and training takes place within that constrained class.

For weakly dissipative dynamics there is a further choice: not only \emph{how} the map is represented, but \emph{what} the learned object is asked to represent. Factoring out an analytically known conservative flow leaves a near-identity interaction-picture map containing only the perturbation. In generator form, \citet{blanco2026magnusian} construct this finite-time correction through the Magnusian expansion. This separation is particularly attractive for radiation reaction, where the dissipative force is parametrically small compared with the conservative orbital dynamics.

The question addressed here is therefore whether a neural flow map can carry a radiation-reaction inspiral, the full sweep from $r_0$ towards coalescence, chained without reinitialization, at an accuracy meaningful next to the integrators built for the job. And what, precisely, does the network contribute beyond the analytic structure on which it is built. 

In this work we test this programme on a controlled compact binary problem as \citet{tsang2015slimplectic, lubich2010symplectic}. We construct and compare symplectic, slimplectic, and Magnusian neural-flow architectures, differing in how much of the finite-time dynamics is supplied analytically and how much is left to be learned. The maps are not assessed only one step at a time: they are composed throughout the inspiral and scored against independent numerical solutions using the accumulated orbital phase and the secular dissipative evolution. This makes the benchmark sensitive precisely to the small structural errors that can remain invisible in short-time prediction.

The aim is threefold: to determine whether structure-preserving neural maps can control secular error while amortizing the cost of repeated inspiral
calculations; to isolate what improvement comes from the learned component
rather than from the analytic structure already built into the architecture;
and to test whether the same secular structure that makes the error controllable also makes an un-modelled dissipative force, like dynamical friction, \citet{chandrasekhar1943dynamical, chandrasekhar1943dynamical_ii, chandrasekhar1943dynamical_iii, ostriker1999dynamical, barausse2014can, kavanagh2020detecting, speri2023probing}, recoverable from the chained phase.

Section~\ref{sec:mechanics} develops the classical machinery: Galley's doubled action, its slimplectic discretization, and the interaction picture, which determines how much of the dynamics is left for a learned object to represent. Section~\ref{sec:neural_flows_architectures} introduces the three learned flow maps \texttt{[SINFONIA-JX]}, their objectives, and the error budget governing whether they can survive $10^3$--$10^6$ compositions. Section~\ref{sec:physics_at_last} specifies the physical benchmark, diagnostics, and results, including a network-off control  \texttt{[SIMFONIA-JX]} for each learned construction. Section~\ref{sec:discussion} summarizes what these experiments establish, and what they do not. Finally we list the incoming work-packages part of the ongoing \texttt{[SIMFONIA$||$SINFONIA]} collaborative research programme \citet{GomesdaSilvasinfoniaWP}. 

\section{Non-conservative mechanics: Discretization and re-organization}\label{sec:mechanics}
\subsection{Galley's doubled action}\label{sec:galley}
Hamilton's principle \citep{goldstein2011classical,hamilton1834general,hamilton1835second} is
commonly formulated as a boundary-value problem: the action is extremized over paths with fixed endpoint data. For dissipative dynamics, by contrast, the natural formulation is an initial-value problem, in which the initial state is specified and the final state is determined by the evolution. Galley's doubled-variable formalism \citep{galley2013classical,galley2014principle} provides a variational principle adapted to this setting by doubling the configuration variables, $q \to (q_1, q_2)$, with the two histories associated with the forward and backward branches of a
closed-time path. Introducing the average and difference variables
\begin{equation}
  \qp = \tfrac{1}{2}(q_1 + q_2), \qquad \qm = q_1 - q_2 ,
\end{equation}
the doubled action may be written as
\begin{equation}
\begin{split}
  S[q_1,q_2] = \int_{t_i}^{t_f}\!\! dt\,\Big[\, &L(q_1,\dot q_1) - L(q_2,\dot q_2) \\
                        &+ K(\qp,\qm,\dot\qp,\dot\qm) \,\Big],
\end{split}
\label{eq:galley-action}
\end{equation}
where $K$ contains the non-conservative interactions and, in the applications considered here, encodes the dissipative dynamics. The two histories are auxiliary variables rather than two independent physical systems. One first varies the doubled action and only afterwards identifies the physical sector through the physical limit (PL),
\begin{align}
  &q_1 = q_2 = q, \quad \pi_1 = -\pi_2 = \pi \nonumber \\
  &\Longleftrightarrow\quad \qm = 0, \quad \pi_+ \equiv \pi_1 + \pi_2 = 0 .
\label{eq:PL}
\end{align}
Here $\pi_{1,2}$ denote the momenta canonically conjugate to $q_{1,2}$ in the doubled Lagrangian, so that the relative minus sign in $L_1 - L_2$ implies $\pi_2 = -p_2$ in the phase-space convention of \citet{blanco2026magnusian}. In the latter convention the physical limit is equivalently $Q_- = 0$, i.e.\ $q_- = 0$ and $p_- = 0$.

The resulting physical equation of motion contains the generalized non-conservative force
\begin{equation}
  \mathcal{Q} =
  \left[ \frac{\partial K}{\partial\qm}
       - \frac{d}{dt}\frac{\partial K}{\partial\dot\qm} \right]_{\rm PL} ,
\label{eq:galley_force}
\end{equation}
which, since the $K$ used below is independent of $\dot\qm$, reduces to $\mathcal{Q} = [\partial K/\partial\qm]_{\rm PL}$.

Three structural features of this construction are important for both the classical
\texttt{[SIMFONIA-JX]} and learned \texttt{[SINFONIA-JX]} maps developed below.

First, the formulation remains variational. Discretizing the action in equation \eqref{eq:galley-action}, rather than the equations of motion directly, produces a non-conservative variational integrator with the corresponding discrete geometric structure and modified Noether evolution laws. This is the basis of the slimplectic construction reviewed in Section~\ref{sec:slim}.

Second, the doubled phase space is eight-dimensional for the planar problem considered here: two copies of the four-dimensional physical phase space $(q,\pi)$. The physical dynamics occupies the four-dimensional PL submanifold of equation \eqref{eq:PL}. At the continuous level, physical initial data select this sector, and the PL surface is preserved by the doubled evolution \citep[his equation 2.14]{blanco2026magnusian}. The explicit PL operation is nevertheless essential in the variational construction: the doubled variables are treated independently during variation and identified only after
the equations of motion have been obtained.

Third, the PL submanifold is Lagrangian with respect to the doubled symplectic form, but the evolution induced on the physical variables is not, in general, symplectic when dissipation is present. Canonical evolution on the doubled phase space must therefore not be confused with canonical evolution, symplecticity, or phase-space-volume preservation of the projected physical dynamics. We return to this distinction in Section~\ref{sec:discussion}.

A further structural result, following equations ~(2.38)--(2.42) of \citet{blanco2026magnusian}, underpins the third architecture, \texttt{[SIMFONIA-J2/SINFONIA-J2]}. Let $Q = (q,p)$ collect the phase-space coordinates, with $A$ running over the four physical directions and $\Omega^{AB}$ the inverse physical symplectic form. Consider a doubled generator linear in the difference variables,
\begin{equation}
  K(Q_+, Q_-) = Q_-^A\, k_A(Q_+) .
\label{eq:linear_doubled_generator}
\end{equation}
Its Poisson action on any physical observable $\mathcal{O}(Q_+)$ is
\begin{equation}
  \{K, \mathcal{O}\} = \Omega^{AB} k_A(Q_+)\, \partial_B\mathcal{O}(Q_+) ,
\label{eq:doubled_vector_field}
\end{equation}
and therefore induces an ordinary vector field on the physical phase space. Moreover, the class in equation \eqref{eq:linear_doubled_generator} is closed under the doubled Poisson bracket, so nested brackets of such generators remain linear in $Q_-$ while their action on physical observables remains independent of $Q_-$. This is precisely the property that allows the dissipative Magnusian to retain the form
\begin{equation}
  \chi(Q_+, Q_-) = Q_-^A\, \chi_A(Q_+)
  \label{eq:magnusian_linear}
\end{equation}
order by order, so that its exponentiated action on physical observables can be evaluated entirely in terms of $Q_+$, without an additional PL projection after each nested bracket.

Conversely, because the physical symplectic form is non-degenerate, the map
\begin{equation}
  k_A(Q_+) \;\longmapsto\; V^B(Q_+) = \Omega^{AB} k_A(Q_+)
  \label{eq:mag_map}
\end{equation}
is invertible. Any smooth vector field $V$ on the physical phase space can therefore be represented by a doubled generator of the form in equation \eqref{eq:linear_doubled_generator}, and hence admits a Hamiltonian lift to the doubled phase space. In this precise sense, requiring the generator to be canonical on the doubled space places no additional restriction on the dissipative physical vector field itself: the canonical structure belongs to its doubled Hamiltonian representation, while the projected physical flow may remain fully non-Hamiltonian and dissipative.

This distinction will be useful below. For \texttt{[SINFONIA-J2]}, learning a vector field directly on $Q_+$ does not discard the doubled-space Magnusian structure: every such learned field admits the corresponding linear-in-$Q_-$ Hamiltonian lift. What is non-trivial in the architecture is therefore not the mere existence of doubled canonicity, but the finite-time generator structure, the interaction-picture factorization, and the additional \underline{physical constraints imposed on the learned map}.

\subsection{The slimplectic integrator}\label{sec:slim}

\citet{tsang2015slimplectic} discretize the doubled action in equation \eqref{eq:galley-action} directly. On each step, $\qp$ and $\qm$ are represented by polynomial interpolants through $\vartheta+2$ Gauss--Lobatto nodes, and the action integral is approximated using Galerkin--Gauss--Lobatto (GGL) quadrature. The resulting discrete action is extremized with respect to the internal degrees of freedom, and only then is the PL in equation \eqref{eq:PL} taken. The outcome is a fixed-step, one-step implicit scheme of order $2\vartheta+2$: a non-conservative variational integrator. We use $\vartheta=1$ throughout, i.e.\ the fourth-order member, \texttt{Slim4}, which serves as the reference integrator against which each architecture is benchmarked.

The term \emph{slimplectic} is due to \citet{tsang2015slimplectic} and denotes the extension of variational integration to non-conservative systems. The resulting maps are not symplectic on the physical phase space, but retain characteristic long-time advantages of variational integrators. In the examples of \citet{tsang2015slimplectic}, this manifests as bounded error in the energy and momenta relative to the true dissipative evolution, together with phase error growing approximately as $\propto t$ rather than the $\propto t^2$ growth exhibited by the corresponding Runge--Kutta schemes. In what follows, we use the term in this operational sense and benchmark these long-time properties, at matched step size,  directly for the inspiral problem considered here.

\subsection{The interaction picture and the Magnusian generator}\label{sec:ip}

The last piece of classical machinery is the one that decides what a numerical  \texttt{[SIMFONIA-JX]} or learned object \texttt{[SINFONIA-JX]} is asked to represent. Writing the vector field as a fast conservative part plus a small perturbation,
\begin{equation}
  \dot z = f_{\rm cons}(z) + \varepsilon\, f_{\rm rr}(z), 
  \qquad \varepsilon \ll 1 ,
  \label{eq:split}
\end{equation}
and letting $\Phi^{\rm cons}_{h}$ denote the exact conservative flow; for the Kepler problem considered here it is available through the universal-variable $f$--$g$ Lagrange coefficients, requiring only the solution of a single scalar Kepler equation. The full map can then be factorized as
\begin{equation}
  \Phi_h(z) \;=\; \Phi^{\rm cons}_{h}\!\big( W_h(z) \big),
  \label{eq:ip}
\end{equation}
with $W_h$ a near-identity map, $W_h = \mathrm{id} + \mathcal{O}(\varepsilon h)$, generated by the perturbation transported into the frame co-moving with the conservative flow. This is the interaction-picture form of the slow--fast decomposition. Closely related constructions appear throughout celestial mechanics and self-force theory: osculating-element methods represent the perturbed motion as a slowly evolving sequence of conservative orbits \citep{brouwer1962methods, pound2008osculating}; Wisdom--Holman splitting \citep{wisdom1991symplectic} realizes the same conservative-flow-plus-perturbation factorization at the level of a symplectic map; and near-identity transformations \citep{van2018fast} and two-timescale expansions \citep{hinderer2008two} systematically separate fast orbital motion from slow secular evolution.

In the doubled-space language of Section~\ref{sec:galley}, the finite-time Magnusian of \citet{blanco2026magnusian} retains the linear form $\chi(Q_+,Q_-) = Q_-^A\,\chi_A(Q_+)$ of equation \eqref{eq:magnusian_linear}. Its coefficients induce the physical-space vector field $V_\chi^{B} = \Omega^{AB}\chi_A(Q_+)$, and the interaction-picture map is the flow of that vector field, $W_h = \exp(V_\chi)$; throughout, we write $\exp(\chi)$ for this flow, the symbol $\chi$ standing for the induced vector field wherever it is exponentiated. Its first-order term is the
perturbing force transported by the Jacobi propagator of the conservative flow, already expressed as the induced physical generator,
\begin{equation}
  \chi^{(1)}(z,h) = \int_0^h M^{-1}(s;z)\, F\!\big(\Phi^{\rm cons}_s(z)\big)\, ds .
  \label{eq:chi1}
\end{equation}
For benchmarking, the exact interaction-picture endpoint is defined pointwise by
\begin{equation}
  W^{\rm true}_h(z) = \Phi^{\rm cons}_{-h}\!\big(\Phi^{\rm true}_h(z)\big)
\end{equation}
\citep[equation 3.7]{blanco2026magnusian}, evaluated by a single one-window integration of the equations of motion per phase-space point: no precomputed trajectory data enters.

At the level of observables, the corresponding Koopman evolution is generated by the Lie action of the same physical vector field, $U_h = \exp(\mathcal{L}_{V_\chi})$; \citet{Canizares:2026tgy} develops this operator-side description for finite-window orbital evolution. Here we instead exponentiate the generator directly on phase space. The operator view has its own literature on this problem, and it is not confined to the relativistic case: \citet{HOFMANN2025314} build analytical (Galerkin) and data-driven Koopman operators by extended dynamic mode decomposition \citep{williams2015data} for Keplerian motion perturbed by oblateness and atmospheric drag, in a regularized set of orbital element, that is, for a dissipatively perturbed two-body problem already written in element variables, while \citet{nehma2025deep} globally linearize the two-body problem itself with a deep auto-encoder in the manner of \citet{lusch2018deep}; the wider programme extends to uncertainty propagation and filtering, control, and transfer design \citep{servadio2023koopman, servadio2024propagation, servadio2025analytical, servadio2025operator}. The operator-side arm of Section~\ref{sec:arch_j2}, tested here as \texttt{[SIMFONIA-J2K||SINFONIA-J2K]}, is of that kind.

Two features distinguish equation \eqref{eq:ip} from the dictionary-based Koopman approximations just cited. First, the analytically known conservative flow is not represented on the dictionary: in the decomposition $K = K_0 + K_1 + \dots$ of \citet[equation ~26]{HOFMANN2025314} the Keplerian term $K_0$ is itself expanded on the basis, whereas equation \eqref{eq:ip} applies $\Phi^{\rm cons}_h$ exactly and leaves only $W_h$ to be approximated. Second, the regime differs by orders: in the low-Earth-orbit benchmarks of \citet{HOFMANN2025314}, orbits are propagated over $\sim\!10^2$ revolutions to \emph{kilometre-level} position error, with the drag force known and supplied to the model. Here the map is composed to coalescence some $10^5$ times and scored on accumulated orbital phase against an independent truth, with the secular channel that governs that phase imposed structurally on each window endpoint rather than inferred from data. In Section~\ref{sec:drag_force} the same channel is instead left free, and a drag force withheld from the model is recovered from the phase it accumulates. \\

This distinction is what makes equation \eqref{eq:ip} a statement about \emph{what is left to represent}. In the benchmark of Section~\ref{sec:physics_at_last} the radiation-reaction force is between four and five orders of magnitude below the conservative one. An approximant asked to represent $\Phi_h$ directly must first resolve the dominant Kepler rotation, while the much smaller dissipative correction can lie below the approximation error of that conservative motion (measured in Section~\ref{sec:arch_j0}): this is the plain flow of Section~\ref{sec:arch_j0}, \texttt{[SIMFONIA-J0||SINFONIA-J0]}, and the Taylor-anchored flow of Section~\ref{sec:arch_j1} is the same choice with the leading orders supplied analytically rather than fitted, \texttt{[SIMFONIA-J1||SINFONIA-J1]}. An approximant asked for $W_h$ sees nothing but the perturbation: the interaction picture variant of Section~\ref{sec:arch_j0} and the operator map of Section~\ref{sec:arch_j2}. And an approximant asked only for $W_h - \exp(\chi^{(1)})$ sees only what no closed form supplies: the residual generator of equation \eqref{eq:j2}, \texttt{[SIMFONIA-J2||SINFONIA-J2]}. These are the three choices the architectures make, and Section~\ref{sec:physics_at_last} measures what each buys.

\section{Neural Flow Maps}\label{sec:neural_flows_architectures}

\subsection{What \enquote{structure-preserving} means for a learned map}\label{sec:gnml}

A learned integrator is a parametric family $\Phi_\theta$ of candidate evolution maps, indexed by the trainable parameters $\theta$, together with an objective that selects one member of that family. Structure can be imposed on the family itself, before any optimization, and three distinct ways of doing so are worth separating. They are often described using similar terminology, despite imposing different constraints on the learned map and behaving differently under repeated composition.

The first is a \emph{geometric} property shared exactly by every member of the family, symplecticity, volume preservation, equivariance under a prescribed symmetry, or reversibility. \texttt{SympNets} \citep{jin2020sympnets} and the symplectic neural flows of \citet{xu2026cosynflow, canizares2024symplectic} belong to this class: the constraint is encoded directly in the map architecture and therefore holds for any $\theta$, independently of training. Hamiltonian neural networks \citep{greydanus2019hamiltonian, sosanya2022dissipative} impose an analogous constraint at the level of the learned vector field, although exact symplecticity of the resulting finite-time map additionally depends on how that vector field is integrated. For Hamiltonian systems, symplectic maps also admit the classical backward-error interpretation: under the usual assumptions, their repeated composition follows the flow of a nearby modified Hamiltonian, leading to bounded, typically oscillatory energy errors over long integrations rather than systematic secular drift \citet{hairer2013geometric}.

The second is an \emph{approximation-order} constraint. The family can be constructed so that every member agrees with a known approximation to the true flow through a prescribed order in the step, with the network entering only at the first unresolved order, the defect-correction strategy of \citet{fang2025learning, fang2025accelerating}. This need not enforce a geometric invariant; rather, it hard-wires the consistency order of the underlying expansion into the learned map.

The third is an \emph{analytic anchor}. The family is organized around a known analytic evolution map, with learning restricted to a bounded residual, so that the analytic contribution remains explicit for every $\theta$ and the network represents only what the baseline omits. This is not an analytic floor: a trained residual can perform worse than the map on which it is built, and in the present experiments sometimes does (Section~\ref{sec:physics_at_last}). What the construction guarantees is instead that the analytic map belongs to the hypothesis class and that departures from it are bounded.

Only the first is structure-preserving in the conventional geometric-integration sense. All three, however, encode prior structure in the hypothesis class rather than asking it to emerge from optimization, and all three are properties of the parametrization itself.

The principal result of this paper concerns a qualitatively different constraint. It is not imposed on the hypothesis class at all, but on the realized map after training: with the network weights frozen, a single scalar constraint derived from the force law is imposed directly on the map output (Section~\ref{sec:arch_jx_secular_channel}). The placement of the known analytic dynamics within the architecture remains important, transferring it from the learned component to the analytic part of the map improves the error by a factor of approximately thirty at the benchmark step (Section~\ref{sec:arch_j0}), but does not determine the long-time outcome. Explicitly controlling the single secular channel of the chained evolution improves the chained error by four orders of magnitude on the benchmark studied here, without retraining the network.

\subsection{The secular channel} \label{sec:arch_jx_secular_channel}

A chained map accumulates orbital phase over many thousands of cycles, so a per-window error $\eta$ chained over $N \gg 1$ windows may accumulate coherently, as $N\eta$, diffusively, as $\sqrt{N}\eta$, or remain much smaller if its signed contribution is periodic and cancels over each orbit. For the quasi-circular inspiral considered here, which regime is realized is controlled by the secular projection of the local map error onto the slow dissipative directions, namely the orbital energy and angular momentum.

On the quasi-circular inspiral tube these two directions are not independent. Circular-orbit flux balance gives
\begin{equation}
\frac{dL_z}{dE}=\frac{1}{\omega},
\label{eq:fluxlock}
\end{equation}
\citep{kennefick1996radiation}, so that energy and angular-momentum loss are locked along the quasi-circular sequence. The dissipative drift therefore contains a single independent secular channel. By contrast, the map's defect in the orbital angle itself over a single window, the phase error it would commit even if the secular $(E, L_z)$ drift were exact, is negligible at the accuracies considered here: the long-time phase error is mediated almost entirely through the secular drift, which sets the future inspiral rate, rather than through the angle directly. Section~\ref{sec:physics_at_last} isolates this single secular projection in five independent diagnostics and then imposes it explicitly; the ability of each architecture to carry the inspiral is determined primarily by how it treats this channel.

To control it, we define $\Pi[\,w;\,E_\star,L_\star\,]$ as the map that carries a window endpoint $w=(\mathbf q,\mathbf p)$ onto the target shell $E=E_\star$, $L_z=L_\star$. Rather than correcting the two invariants independently, the projection acts along the epicyclic degree of freedom transverse to the quasi-circular sequence. Writing $r=|\mathbf q|$, $\hat{\mathbf q}=\mathbf q/r$, $p_r=\mathbf p\cdot\hat{\mathbf q}$, and $r_c=L_\star^2$ for the circular-orbit radius associated with the target angular momentum, we rescale the epicyclic excursion (the deviation $(r-r_c,\;p_r)$ from that circular orbit) according to
\begin{equation}
(r-r_c,\;p_r)\ \longmapsto\ \lambda\,(r-r_c,\;p_r),
\qquad
p_\theta=\frac{L_\star}{r'},
\label{eq:proj}
\end{equation}
where $r' = r_c + \lambda\,(r-r_c)$ is the rescaled radius. The non-negative scalar $\lambda$ is determined from $E(\lambda)=E_\star$; in practice, four Newton iterations are sufficient.

The targets themselves are obtained entirely from the force law,
\begin{equation}
E_\star=E(z)+\Delta E,
\qquad
L_\star=L_z(z)+\Delta L_z,
\end{equation}
where $(\Delta E,\Delta L_z)$ are the force-law fluxes integrated along the conservative arc over the window. No reference trajectory or independently solved inspiral is required. Because the correction rescales the epicyclic excursion while adjusting the tangential momentum consistently, it moves $\mathbf q$ and $\mathbf p$ together and satisfies both target invariants to the projection tolerance. Simpler alternatives are ill-conditioned on the same quasi-circular tube. Projection onto an invariant shell has a classical lineage, the manifold corrections of \citet{fukushima2003efficient, zhong2010manifold} \citep[chapter IV.4]{hairer2013geometric}, and, closest to the present construction, \citet{luo2024dissipated} project a post-Newtonian binary onto the energy shell predicted by its own dissipated energy at each step. The projection used here differs in the joint $(E, L_z)$ target enforced through the flux lock and in the window, one to three orders of magnitude coarser; the fuller comparison is given in Section~\ref{sec:discussion}. Holding $\mathbf q$ fixed, for example, transfers the radial discrepancy entirely into $p_r$ through a near-cancellation and fails once $\delta r\gtrsim p_r^2$, which occurs for windows longer than $\sim\tisco$ in the present scaling. A minimal-norm correction likewise becomes singular in the quasi-circular limit as the constraint Jacobian loses rank, $\nabla E\parallel\nabla L_z$.

The mismatch between one-step training objectives and autoregressive rollout is well documented across learned dynamical systems and sequence prediction
\citep{ross2011reduction, bengio2015scheduled, sanchez2020learning, brandstetter2022message, lam2023learning, krishnapriyan2021characterizing}. Recent work has further cautioned against interpreting correlations between local and rollout error causally \citep{Thummler:2026baz}, and has shown that long-time error can sometimes be localized to a small number of physical channels and corrected after training \citep{samanta2026autoregressive}. We take that caution seriously and rely on controlled comparisons at fixed pointwise error rather than on correlations across models. The contribution here is more specific: the relevant channel is derived directly from the orbital balance law rather than inferred from rollout data, and its signed projection explains the ordering of the learned maps.

Finally, \underline{and by design}, none of the maps considered here is trained on a precomputed trajectory library. There is no catalogue of reference inspirals: the training signal is generated directly from the governing equations, while high-accuracy trajectories are used only for validation and benchmarking. This distinction is potentially important beyond the present proof of concept. We return to its implications in Section~\ref{sec:discussion}.

\subsection{Symplectic and slimplectic neural flows: \texttt{[SINFONIA-J0]}}
\label{sec:arch_j0}

The first architecture, \texttt{[SINFONIA-J0]}, learns the finite-time flow map directly on Galley's doubled phase space. A symplectic neural flow is constructed as a composition of layers, each of which is the exact time-$t$ flow of a learned separable, time-dependent Hamiltonian; the resulting map is therefore exactly symplectic for arbitrary network parameters. Applied to the eight-dimensional doubled state $(\qp,\qm,\pi_+,\pi_-)$, with the physical limit of equation \eqref{eq:PL} imposed only after evaluation of the map, this provides the learned analogue of the slimplectic construction: evolution is symplectic on the doubled space, while the induced physical dynamics is dissipative.

We construct two variants. The plain architecture asks the neural layers to represent the complete doubled vector field equation \eqref{eq:split}. The interaction-picture variant instead embeds the exact Kepler flow into the architecture through equation \eqref{eq:ip} and asks the neural component only to represent the near-identity factor $W_h$, whose deviation from the identity is $\mathcal{O}(\rr h)$. Both variants are trained from the physics-informed residual
\begin{equation}
\left|\partial_t\Phi_\theta-f_{\rm doubled}(\Phi_\theta)\right|,
\end{equation}
evaluated at collocation points in a neighbourhood of the physical-limit surface. In the interaction picture this residual is divided by $\rr$, so that the optimization problem remains $\mathcal{O}(1)$. We use a Huber loss rather than a squared residual throughout. 

The failure of the plain \texttt{J0} map is quantitative. Its one-window error scales linearly with the window duration, corresponding to zeroth-order consistency, and closely follows the flow of a slightly incorrect vector field with $|\delta f|\simeq5\times10^{-3}$. The approximation error associated with the dominant $\mathcal{O}(1)$ Kepler motion surpasses the scale of the $\mathcal{O}(\rr)$ dissipative contribution by about a factor of $300$ at $r_0$ and remains ${\sim}20$ times larger at the end of the scored interval. Within this unit-slope regime, reducing the chaining window is not an effective lever: the error per window scales as $h$ while the number of windows scales as $h^{-1}$, leaving the coherently accumulated error approximately independent of $h$. By contrast, embedding the Kepler flow exactly changes only what the network is required to represent and improves the resulting map by approximately a factor of $30$ at fixed sampler, timestep, training horizon, and optimization protocol.

This comparison also separates two effects that would otherwise be conflated. Both neural constructions are symplectic on Galley's doubled space by design; doubled-space symplecticity therefore cannot explain the difference between them. The improvement instead arises from the interaction-picture decomposition, which removes the dominant conservative motion before the smaller dissipative dynamics is learned.

The same principle does not require the dominant flow to be available analytically. For higher-order post-Newtonian models, generic spin dynamics, or effective-one-body systems in which an exact conservative finite-time flow is unavailable, the inner factor in equation \eqref{eq:ip} is itself represented by a symplectic neural flow. The construction is therefore applied hierarchically: a structure-preserving approximation carries the dominant conservative dynamics, while successive maps learn progressively smaller corrections. The relevant requirement is quantitative rather than formal, the error of the inner map must remain below the scale of the perturbation delegated to the outer map. This provides a directly measurable criterion for when such a recursive decomposition is useful. Further results are to be featured in \citet{Simfonia_Sinfonia_PN_EOB}. In the interim we encourage the reader to have a look at preliminary results before significant architectural refinements in \citet{GomesdaSilva2026sinfonia}. 

To clarify on the nomenclature, for the present benchmark, however, \texttt{[SINFONIA-J0]} denotes only the learned construction introduced above. Unlike the architectures of Sections~\ref{sec:arch_j1} and \ref{sec:arch_j2}, it has no network-free counterpart of its own in our experiments: the classical baseline against which it is scored is the fourth-order slimplectic integrator \texttt{Slim4} of \citet{tsang2015slimplectic}, in the authors' implementation \citep{slimplecticcode}, which is not a \texttt{J} arm on my architecture in this publication. \footnote{We have additionally developed an optimized \texttt{JAX} implementation of \texttt{Slim4} and an adaptive variant. Neither is expected to alter the separation from the learned maps of Section~\ref{sec:physics_at_last} at matched accuracy, nor speed, and a systematic comparison is deferred to future work; the implementations are available from the author on request.}

\subsection{A Taylor-anchored neural flow: \texttt{[SIMFONIA-J1/SINFONIA-J1]}}\label{sec:arch_j1}

The second architecture anchors the map on the Taylor expansion of the full field and learns the remainder \citep{fang2025learning, fang2025accelerating}:
\begin{equation}
  \Phi_\theta(u,h) \;=\; u + \sum_{k=1}^{p}\frac{h^k}{k!}f_k(u)
  \;+\; \mathcal{A}(u,h)\,\Delta_\theta(u,h),
  \label{eq:j1}
\end{equation}
with $f_1 = f$, $f_{k+1} = (\partial f_k/\partial u)\,f$ the iterated Lie derivatives of the vector field (evaluated by Taylor-mode automatic differentiation; a fixed, non-learnable tower), $p=5$, and $\Delta_\theta$ an equivariant multilayer perceptron reading rotational invariants of the state. 
Two choices distinguish the present form from \citet{fang2025learning, fang2025accelerating}. The amplitude prior $\mathcal{A} = h^{6}\|f_6\|/720$ is the size of the first omitted Taylor term, so the network emits an $\mathcal{O}(1)$ shape function (a bare $h^{p+1}$ gate at $h\le 10^{-2}$ puts the gradient eleven orders below the optimizer's numerical floor, and the remainder never trains; see Section~\ref{sec:discussion}); and the map's endpoint is projected onto the energy--angular-momentum shell that the force law requires over the window, computed by Gauss--Legendre quadrature of the force-law fluxes along the map's own window arc. We also omit the time activations of \citet{fang2025learning} (their equations 27--29), whose role over the fixed windows used here is played by the amplitude prior. Three key features hold for any weights: \texttt{[F1]:} exact consistency ($\Phi_\theta(u,0)=u$), \texttt{[F2]:} order confinement (the learned term enters at $h^{6}$, so training can only choose the coefficient of the leading unresolved term), and \texttt{[F3]:} exact secular budget ($\Delta E$, $\Delta L_z$ per window are the force law's, not the network's).

\texttt{J1}'s working objective is \textbf{unsupervised}: the ODE residual of the map itself, $\|\partial_h\Phi_\theta(u,h) - f(\Phi_\theta(u,h))\|^2$, with $\partial_h$ taken by forward-mode differentiation through the whole map and the force evaluated along the map, sampled at $h\ge\tisco/4$ and normalized by the amplitude prior. A supervised alternative, distillation onto the exact one-window endpoint from a \texttt{DOP853} solve of the force law per sample, reaches the same tier with a wider seed band. Training is $2\times10^{4}$ Adam steps; the shell projection is applied \emph{at deployment, outside both training objectives}, so the loss never sees the secular channel.

\texttt{J1} can carry the inspiral swiftly, for two reasons that are separately measurable. Order confinement makes the learned error small ($\mathcal{O}(h^6)$ against an $\mathcal{O}(h^5)$ tower); the shell projection makes its secular projection \emph{zero} by construction, so what remains of the network's error is exactly the part that cancels around the orbit. With the projection off the same trained remainder lands at $10^{-3}$--$10^{-1}$~rad; with it on, $10^{-6}$. The projection alone ($\Delta_\theta\equiv0$ on the order-5 tower) gives $1.35\times10^{-4}$; the order-7 tower alone with the projection, below $10^{-7}$ (truth-limited). The network's contribution is therefore bounded on both sides by analytic maps of the same family, and it is measured against them.

\begin{figure*}[t]
\centering
\includegraphics[width=0.96\textwidth]{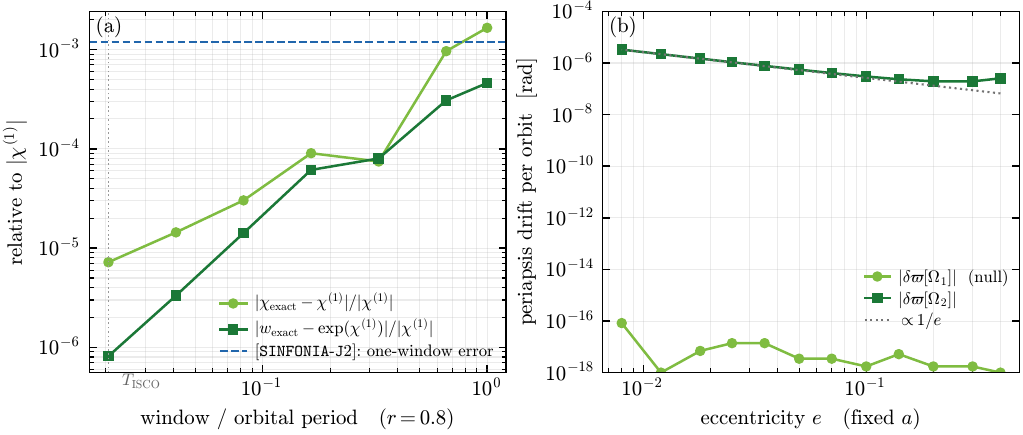}
\caption{(a) \emph{How much Magnusian is there to learn:} Relative to $|\chione|$, at $r=0.8$: the single-bracket residual $|W^{\rm exact}-z-\chione|$ (slope $\approx1$) and the genuine higher-order content $|W^{\rm exact}-\exp(\chione)|$ (slope $\approx2$), versus window length, with the trained networks' fit floor. Below one orbit a network cannot add accuracy to the exponentiated closed form; at one orbit it can. (b) \emph{What the second-order contains:} The same generator at fixed semi-major axis, swept in eccentricity: the per-orbit drift $\delta\varpi[\Omega] = \nabla\varpi(z)\cdot\Omega$ it carries in the periapsis angle $\varpi = \arg\!\big[(v^2-1/r)\mathbf q - (\mathbf q\cdot\mathbf v)\mathbf v\big]$, the argument of the Laplace--Runge--Lenz vector and the Cartesian stand-in for the Delaunay $g$ whose conjugate is $\chi_G$. At first order it vanishes to machine precision at every $e$, at second order it does not, and follows a $1/e$ asymptote (dotted) on the small-$e$ branch. That pole is a property of the chart, not the dynamics: $\varpi$ is the argument of a vector of length $e$, so $|\nabla\varpi|\sim1/e$, while $|\Omega_2|$ itself converges to $3.9\times10^{-7}$ as $e\to0$. The lower curve is floating-point noise, not a measurement.}
\label{fig:content}
\end{figure*}

\subsection{The Magnusian neural flow: \texttt{[SIMFONIA-J2(/K)/SINFONIA-J2(/K)]}}\label{sec:arch_j2}

The third architecture learns the finite-time generator of \eqref{eq:ip} itself:
\begin{equation}
  \Phi_\theta(z) = \Phi^{K}_{\Delta t}\!\Big(\exp\big(V_\theta\big)(z)\Big),\quad
  V_\theta = \chione + \mathcal{A}(z,\Delta t)\,\rho_\theta(z) 
  \label{eq:j2}
\end{equation}
on the four-dimensional physical phase space, no doubling and no physical-limit projection, by \citet{blanco2026magnusian}'s $\qm$-independence. $\Phi^K$ is the exact universal-variable Kepler flow; $\chione$ is given by equation  \eqref{eq:chi1} evaluated by automatic differentiation of $\Phi^K$ (Gauss--Legendre quadrature, machine-exact at 4--16 nodes); $\rho_\theta$ is an equivariant residual field assembled in the $(\hat q,\hat\theta)$ frame from rotational invariants (one $64\times3$ \texttt{tanh} network); $\mathcal{A}$ is the measured scaling of the higher-order Magnus content $\chitwo$ in the slow variable $a=-1/2E$, clipped; and $\exp$ is a converged implicit midpoint, equal to the time-1 flow of $V_\theta$ to $10^{-10}$. The map is exactly equivariant, a genuine flow (reversible to machine precision), and at zero weights it is the exponentiated analytic Magnusian, with the clipped prior $\mathcal{A}$ bounding how far any weights can move it: the analytic map is in the family and the deviation is bounded, though a trained residual can still sit above it (Section~\ref{sec:physics_at_last}). Its window is one orbital period, $\Delta t = \torb(z) = 2\pi a^{3/2}$, the window of \citet{blanco2026magnusian}'s own construction and \emph{the one at which the residual has the most to carry}: the Magnus content beyond $\chione$, measured in Figure~\ref{fig:content}a. Its dominant piece, the second-order term $\Omega_2$, we evaluate numerically below, its closed form deferred to \citep{Magnusian2}, and the content beyond $\Omega_2$ has no closed form at all.

We evaluate that second-order term rather than assume it. In Magnus notation, with $A(s) = M^{-1}(s;z)\,F\!\big(\Phi^{\rm cons}_s(z)\big)$ the integrand of equation \eqref{eq:chi1}, the first term is $\Omega_1 = \int_0^{T}\!A(s)\,ds = \chione$ and the second is $\Omega_2 = \tfrac12\!\int_0^T\!ds_1\!\int_0^{s_1}\!ds_2\,\big[A(s_2),A(s_1)\big]$, with $[\cdot,\cdot]$ the commutator of vector fields. $\Omega_2$ is computed by nested Gauss--Legendre quadrature with the Jacobians of the interaction-picture field by automatic differentiation, the bracket sign fixed empirically against the exact endpoint (the one-window residual falls $10^{2}$--$10^{4}$ times with one sign and doubles with the other).

Two properties of the result follow, both of the generator alone (Figure~\ref{fig:content}b). Swept in eccentricity at fixed semi-major axis, the component conjugate to the angular momentum, the one driven by the drift of the periapsis angle, vanishes at first order to $|\delta\varpi[\Omega_1]|/|\Omega_1| \le 10^{-14}$ at every eccentricity, reproducing in Cartesian
interaction-picture variables the null $\chi_G = 0$ that \citet[Appendix D]{Canizares:2026tgy} validates in Delaunay variables against \citet{blanco2026magnusian}'s closed form. At second order it does not vanish: it is ten orders above that null and converged to $10^{-13}$ under refinement of the nested quadrature. The second order therefore reaches a direction the first order cannot at any amplitude, which is why the second-order base improves on $\exp(\chione)$ rather than merely rescaling it. That projection grows as $e^{-0.96}$ on the branch $e\le0.1$ while $|\Omega_2|$ itself converges, to $3.9\times10^{-7}$ as $e\to0$: the singularity is in the chart, not the dynamics. The periapsis angle is the argument of a vector of length $e$, so $|\nabla\varpi|\sim1/e$ and any element-based representation inherits the pole, while the Cartesian covector field of \eqref{eq:j2} never forms $\varpi$ and does not. The closed form of the pole in Delaunay variables is derived in \citep{Magnusian2}.

An earlier form of this architecture, the whole generator learned, a
Helmholtz split $J(\nabla S+\rho)$, deployed at Tsang's fixed step, appears in the results as the unconstrained and shell-projected \texttt{J2} arms; its lessons are in Section~\ref{sec:discussion}.
\begin{figure*}[t]
\centering
\includegraphics[width=0.86\textwidth]{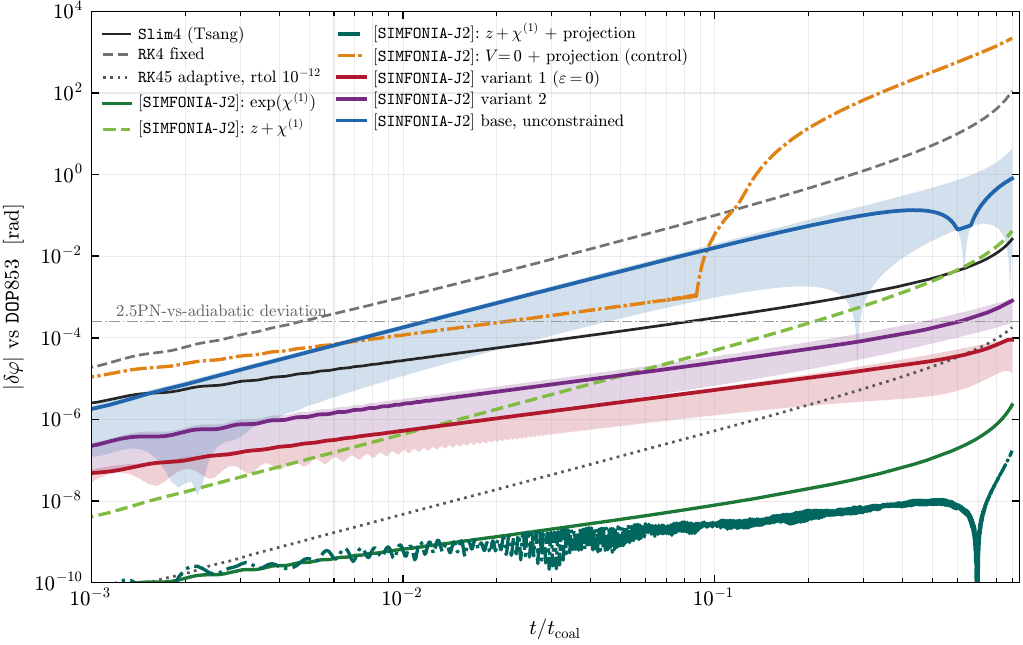}
\caption{Accumulated phase error against the \texttt{DOP853} truth for the classical arms (\texttt{Slim4}, \texttt{RK4} at the same fixed-step, adaptive \texttt{RK45}), the network-free Magnusian maps (single bracket $z+\chione$; exponentiated $\exp(\chione)$; $z+\chione$ with the secular shell projection; the $V=0$ control), and the earlier \texttt{J2} learned arms (three seeds each, median and band). The horizontal grey line marks the physical deviation between the full 2.5PN dynamics and its leading-order adiabatic approximation ($2.5\times10^{-4}$~rad at $0.9\ t_{\rm coal}$): a reference scale, not a numerical floor, an arm below it is closer to the 2.5PN truth than the adiabatic description of the same binary, and an error quoted against the adiabatic formula could not resolve anything beneath it. The exponentiated analytic map is the strongest integrator at this step; the shell projection removes the fixed-sign residue of whatever generator it is given. }
\label{fig:tisco}
\end{figure*}
\paragraph{The same generator on the operator side.} Equation \eqref{eq:j2} learns the log of the interaction-picture map. The Koopman reading of Section~\ref{sec:ip} suggests learning the \emph{exp} instead, and we include it as a representation check rather than a fourth architecture, since it is the same object and the same three choices of Section~\ref{sec:gnml}. Writing
\begin{equation}
  \Phi_\theta(z) = \Phi^{K}_{\Delta t}\big( \mathcal{D}_z[\,A\,\psi(x(z))\,] \big),
  \label{eq:j3}
\end{equation}
with $x(z)=(r,\sqrt r\,p_r,\sqrt r\,p_\theta)$ the three independent rotational invariants, $\psi$ a tensor--Chebyshev dictionary in $x$ up to total degree $d$, $A$ a $4\times N_\psi$ matrix obtained in closed form by singular-value least squares against the same force-law-only target
$W^{\rm exact}_{\Delta t}$, and $\mathcal{D}_z$ the decoder that returns the endpoint in $z$'s own orbital frame, with an amplitude prior of the same measured scaling as in equation \eqref{eq:j2}. There is no optimizer, no learning rate and no initialization: the only stochasticity is the data draw. The shell projection is applied exactly as before. We call this the Koopman side for the reason given in Section~\ref{sec:ip}, with one caveat: no finite polynomial dictionary is closed under the flow of a chirp, so equation \eqref{eq:j3} is extended dynamic mode decomposition with a decoder, the operator restricted to the observables one actually needs, as in \citet{HOFMANN2025314}. \emph{\underline{The asymmetry matters beyond this problem}:} $\chione$ requires only a differentiable conservative flow, its Jacobi propagator and a force evaluation, all available for an effective-one-body model, where no closed-form averaging integral exists, whereas the operator route requires a dictionary in which the dynamics is approximately closed, and the strong-field case is where such a dictionary is hardest to find. We call equation's \eqref{eq:j3} architecture \texttt{[SIMFONIA-J2K]}, the classical member of this pair; its learned counterparts, the same dictionary fitted by Adam in place of least squares, and an MLP encoder in place of $\psi$, are
\texttt{[SINFONIA-J2K]}. Section~\ref{sec:physics_at_last} reports all three, and the difference between them is the point.

\texttt{J2} is distilled onto the exact interaction-picture endpoint $W^{\rm true}_{\Delta t}(z) = \Phi^K_{-\Delta t}(\Phi^{\rm true}_{\Delta t}(z))$ of Section~\ref{sec:ip}, converged force-law targets, no solved trajectory, with the loss $\|(\exp(V_\theta)(z) - W^{\rm true}_{\Delta t}(z))/\mathcal{A}\|^2$ on the model's own exponential, so no Magnus truncation enters the teacher. A second, secular-aware term is applied as a $10$k-step fine-tune: the squared, phase-weighted, signed mean of the relative $(\Delta E,\Delta L_z)$ defect over a fixed quasi-circular panel with exact targets, expressed in predicted accumulated radians. It penalizes only the part of the residual error that accumulates. Both are force-law-only; neither uses a trajectory. \texttt{J2} can, by an argument in two steps. At Tsang's step ($1.5\%$ of an orbit) the exact generator differs from $\exp(\chione)$ by $2\times10^{-7}$--$8\times10^{-5}$ of $|\chione|$ across the inspiral (Figure~\ref{fig:content}); a \texttt{tanh} network fits a residual to $\sim10^{-3}$ of its prior; so at that window the residual architecture can add nothing, and the exponentiated closed form is the map to use,  which it is, at $2.3\times10^{-6}$~rad. One orbit per step the $\chitwo$ content is $10^{-3}$ to a few times $10^{-3}$ of $\chione$, growing further towards small separation, growing with the window as the Magnus scaling predicts, and it has no closed form: \emph{the network has something to learn, the analytic part bounds how much harm it can do, and the secular-aware term pins the one direction the pointwise objective cannot see.}

\begin{deluxetable*}{lrrcrc}
\tablecaption{Orbital-phase error $\Delta\phi$ relative to the \texttt{DOP853} reference solution at $0.9\ t_{\rm coal}$. Reported values are medians over the indicated number of random seeds, with the corresponding spread; all methods satisfy ${\rm trk}@10\%=100\%$. Computational cost is reported in $\mu{\rm s}$ per window on a single CPU core, using the median of five runs. \texttt{Slim4} and \texttt{RK4} are implemented as \texttt{NumPy} loops, whereas the learned and analytic maps use \texttt{JIT}-compiled \texttt{JAX} scans; the resulting timing values are therefore implementation dependent, while the accuracy comparisons are not. The final column gives the log--log growth exponent of $\Delta\phi(t)$ measured over $0.02$--$0.6\,t_{\rm coal}$; the slimplectic long-time criterion of \citet{tsang2015slimplectic} corresponds to approximately linear growth, $\Delta\phi\propto t$. Following Section~\ref{sec:neural_flows_architectures}, \texttt{[SIMFONIA]} denotes the network-free member of an architecture and \texttt{[SINFONIA]} its learned counterpart.\label{tab:main}}
\tablehead{\colhead{arm} & \colhead{rad} & \colhead{seeds / band} & \colhead{net off} &
           \colhead{$\mu$s} & \colhead{exp.}}
\startdata
\multicolumn{6}{l}{\emph{$\Delta t = \tisco$, $76{,}143$ windows}} \\
\texttt{RK4} fixed                                          & $113$                 & ---       & --- & 71  & 2.4 \\
\texttt{Slim4} \citep{tsang2015slimplectic}                            & $2.6\times10^{-2}$    & ---       & --- & 489 & 1.3 \\
{[}\texttt{SIMFONIA-J2}]: $z+\chione$ (single bracket) & $4.1\times10^{-2}$    & ---       & --- & 12  & 2.4 \\
{[}\texttt{SINFONIA-J2}]: neural Magnusian, unconstrained & $0.81$                & 3, $366\times$ & --- & 15 & 1.8 \\
{[}\texttt{SINFONIA-J2}]: + secular shell projection & $8.8\times10^{-5}$    & 3, $7\times$   & --- & 20 & \textbf{1.0} \\
\emph{physical 2.5PN-vs-adiabatic deviation}       & $2.5\times10^{-4}$    &           &     &     &     \\
\texttt{RK45} adaptive, rtol $10^{-12}$                     & $1.8\times10^{-4}$    & ---       & --- & $3\!\times\!10^{6}$ evals & 2.2 \\
{[}\texttt{SIMFONIA-J2}]: $\exp(\chione)$ & $2.3\times10^{-6}$    & ---       & --- & 40  & 1.8 \\
{[}\texttt{SIMFONIA-J2}]: $z+\chione$ + shell projection & $1.7\times10^{-7}$    & ---       & --- & 15  & 2.4 \\
{[}\texttt{SIMFONIA-J1}]: Taylor-7 + shell projection & $\mathbf{<10^{-7}}$   & ---       & $6.6\times10^{-3}$ & \textbf{8.8} & --- \\
{[}\texttt{SINFONIA-J1}]: Taylor-5 + remainder + projection, ODE residual & $3.1\times10^{-6}$ & 3, $4.4\times$ & $1.35\times10^{-4}$ & 14 & --- \\
\quad same, \texttt{DOP853} distillation & $6.6\times10^{-6}$    & 3, $10.5\times$ & $1.35\times10^{-4}$ & 14 & --- \\
{[}\texttt{SINFONIA-J2}]: residual net + $L_z$ projection & $5.6\times10^{-7}$    & 3, $14\times$ & $2.3\times10^{-6}$ & 71 & --- \\
{[}\texttt{SIMFONIA-J2K}]: Koopman/EDMD $d=8$, least squares, raw & $1.07\times10^{-2}$   & 3, $2.3\times$ & --- & 4.2 & --- \\
\quad + shell projection                           & $6.1\times10^{-7}$ & 3, $1.0\times$ & $1.07\times10^{-2}$ & 6.8 & --- \\
{[}\texttt{SINFONIA-J2K}]: Koopman, Adam on the same dictionary & $1.4\times10^{-5}$    & 3, $3.0\times$ & $6.1\times10^{-7}$ & 6.8 & --- \\
{[}\texttt{SINFONIA-J2K}]: Koopman, MLP encoder & $8.1\times10^{-6}$    & 3, $5.2\times$ & $6.1\times10^{-7}$ & 5.2 & --- \\
\hline
\multicolumn{6}{l}{\emph{one map per orbit, $\Delta t = \torb(z)$, $1{,}518$ windows}} \\
{[}\texttt{SIMFONIA-J2}]: $\exp(\chione)$ & $2.85\times10^{-2}$   & ---       & --- & 454 & 1.9 \\
{[}\texttt{SIMFONIA-J2}]: $\exp(\Omega_1+\Omega_2)$ & $1.51\times10^{-2}$   & ---       & --- & 1572 & --- \\
{[}\texttt{SINFONIA-J2}]: residual net on $\Omega_1+\Omega_2$ & $2.8\times10^{-3}$    & 3, $1.25\times$ & $1.51\times10^{-2}$ & 1163 & --- \\
\quad + secular-aware objective                    & $\mathbf{7.0\times10^{-4}}$ & 3, $19\times$ & $1.51\times10^{-2}$ & 1187 & --- \\
\quad\quad best seed                               & $6.4\times10^{-5}$    &           &     &     &     \\
{[}\texttt{SINFONIA-J2}]: residual net & $2.2\times10^{-2}$    & 5, $2.7\times$ & $2.85\times10^{-2}$ & 143 & 1.2 \\
\quad + secular-aware objective                    & $\mathbf{1.0\times10^{-2}}$ & 5, $24\times$ & $2.85\times10^{-2}$ & 136 & 2.1 \\
\quad\quad best seed                               & $1.6\times10^{-3}$    &           &     &     &     \\
\hline
\multicolumn{6}{l}{\emph{one network trained across windows $\tisco$--$4 ,\torb$, see (Figure~\ref{fig:ladder})}} \\
\quad $\Delta t = 4\ \torb(z)$, $379$ windows              & $\mathbf{1.26\times10^{-1}}$ & 3, $1.06\times$ & $7.06$ & 375 & --- \\
\quad $\Delta t = 2\,\torb(z)$, $759$ windows              & $3.3\times10^{-2}$    & 3, $35\times$  & $0.44$ & 426 & --- \\
\quad $\Delta t = \torb(z)$, $1{,}518$ windows             & $0.10$                & 3, $1.2\times$ & $2.85\times10^{-2}$ & 410 & --- \\
\quad $\Delta t = \tisco$, $76{,}143$ windows              & $2.4\times10^{-5}$    & 3, $2.4\times$ & $2.3\times10^{-6}$ & 376 & --- \\
\enddata
\end{deluxetable*}
\section{Example: Gravitational radiation reaction and a hidden environment}\label{sec:physics_at_last}
\subsection{The radiation reaction inspiral}\label{sec:grr_toy}
\subsubsection{The physical system}\label{sec:system}
In this work we take \citet{tsang2015slimplectic}'s 2.5PN toy model without any modification. Two $1.4\ \Msun$ neutron stars, symmetric mass-ratio $\nu = 1/4$, $G = c = 1$, in a planar relative orbit at initial separation $r_0 = 100\ M$. The conservative sector is Newtonian, and the dissipation is the leading-order 2.5PN radiation-reaction force \citep{burke1971gravitational, peters1964gravitational}, order-reduced, entering \eqref{eq:galley-action} through
\begin{align}
K ={}&
\frac{16}{5}\nu^2 M^4
\frac{\dot{\mathbf{q}}_{+}\cdot\dot{\mathbf{q}}_{-}}
     {|\mathbf{q}_{+}|^4}
-\frac{48}{5}\nu^2 M^3
\frac{|\dot{\mathbf{q}}_{+}|^2
      (\dot{\mathbf{q}}_{+}\cdot\mathbf{q}_{-})}
     {|\mathbf{q}_{+}|^3}
\nonumber\\
&+24\nu^2 M^3
\frac{(\dot{\mathbf{q}}_{+}\cdot\mathbf{q}_{+})^2
      (\dot{\mathbf{q}}_{+}\cdot\mathbf{q}_{-})}
     {|\mathbf{q}_{+}|^5}
\nonumber\\
&+\frac{16}{15}\nu^2 M^4
\frac{(\dot{\mathbf{q}}_{+}\cdot\mathbf{q}_{+})
      (\mathbf{q}_{+}\cdot\mathbf{q}_{-})}
     {|\mathbf{q}_{+}|^6}
\nonumber\\
&+\frac{144}{5}\nu^2 M^3
\frac{|\dot{\mathbf{q}}_{+}|^2
      (\dot{\mathbf{q}}_{+}\cdot\mathbf{q}_{+})
      (\mathbf{q}_{+}\cdot\mathbf{q}_{-})}
     {|\mathbf{q}_{+}|^5}
\nonumber\\
&-40\nu^2 M^3
\frac{(\dot{\mathbf{q}}_{+}\cdot\mathbf{q}_{+})^3
      (\mathbf{q}_{+}\cdot\mathbf{q}_{-})}  
     {|\mathbf{q}_{+}|^7},
\label{eq:K}
\end{align}
in the non-dimensionalization $M = \mu = r_0 = 1$, in which the entire strength of radiation reaction collapses to a single scalar,
\begin{equation}
  \rr = \nu\Big(\frac{M}{r_0}\Big)^{5/2} = 2.5\times10^{-6},
  \qquad
  \tcoal = \frac{5}{256\,\rr} = 7812.5 .
  \label{eq:rr}
\end{equation}
Integrating to $0.9 \ \tcoal$ takes the binary from $r = 100\ M$ to $r = 56\ M$ through $1518$ orbits and $9536$~rad of accumulated orbital phase. The system carries no orbital resonances at any order,  every error below is a propagator error. Equation \eqref{eq:rr} organizes everything else, though $\rr$ is the coupling and not itself the force ratio, since the coefficients of equation \eqref{eq:K} multiply it: measured against $|f_{\rm cons}| = M/r^{2}$, the reaction force is $1.6\times10^{-5}$ of it at $r_0$ and $6.8\times10^{-5}$ at $r = 56\ M$. \emph{That is, radiation reaction is four to five orders of magnitude weaker than the conservative force across the scored window.}

At $10^5$ chained windows, diagnostic errors can readily dominate the physical propagation error. We identified five reproducible diagnostic failure modes, each of which produced a plausible but incorrect chained phase error during this work: \emph{unwrap aliasing} (unwrapping the phase on a grid so coarse that the advance between samples exceeds $\pi$, silently dropping whole cycles); \emph{nearest-neighbour time matching} (pairing two arms' outputs by closest time when they live on different grids, at $\omega \simeq 2.4$ a mismatch of half a step is ${\sim}1$~rad); \emph{flat extrapolation} (an interpolant clamping to its last value beyond an arm's grid, freezing that arm's phase while the truth advances); \emph{decimated scoring} (measuring the error on a grid thinned for plotting, which imposes its own floor); and \emph{a phase quoted after the orbit stopped tracking} (a small $\dphi$ read off a trajectory that has already left the true orbit, which the ${\rm trk}$ column below exists to catch).
\begin{figure*}[t]
\centering
\includegraphics[width=0.96\textwidth]{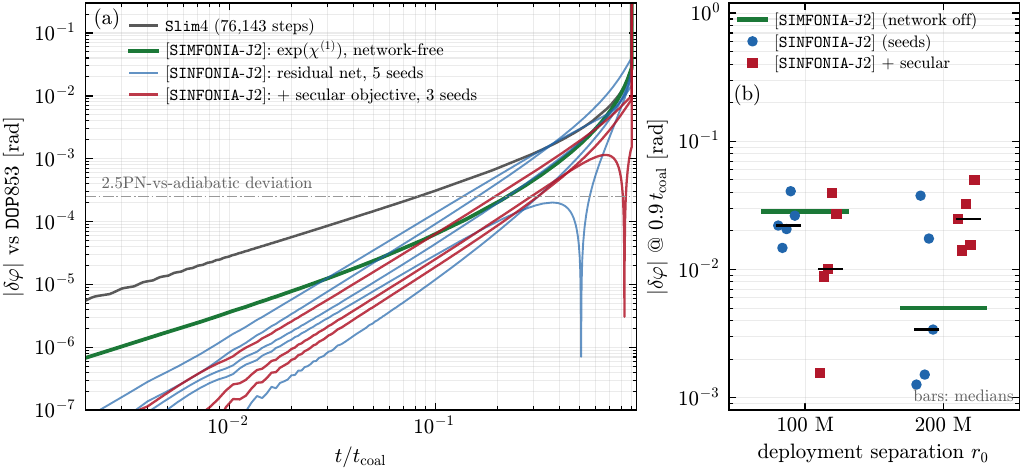}
\caption{Performance of the learned Magnusian with one map evaluation per orbit. Panel (a) shows the results at the training radius, $r_0=100 M$, for five plain \texttt{SINFONIA-J2} seeds, three secular-tuned seeds, the network-off Magnusian control $\exp(\chi^{(1)})$, and the fourth-order slimplectic reference \texttt{Slim4}, evaluated with $76{,}143$ timesteps. Panel (b) shows out-of-distribution deployment of the same networks, trained only at $r_0=100 M$ and at $r_0=200M$ without retraining or modification. Bars denote medians over seeds, the solid line marks the network-off control.}
\label{fig:perorbit}
\end{figure*}
The scoring protocol is therefore fixed once and applied to every arm. All scores are computed against a \texttt{DOP853} truth, never the adiabatic formula, which is itself
$2.5\times10^{-4}$~rad off the 2.5PN dynamics at $0.9\ \tcoal$, on each arm's full undecimated grid, interpolated onto the truth grid with extrapolation masked. Every learned row carries a tracking length ${\rm trk}@10\%$ (the fraction of the inspiral integrated before $|\delta r/r|$ exceeds $10\%$), to be read before the phase column, and is printed beside the identical map with the network switched off: \emph{that column, not the classical references, is what decides what the network adds.}

Finally, the reference's own error bounds what can be claimed. Solved at rtol $=10^{-12}$ the truth is wrong by $1.5\times10^{-6}$~rad at $0.9\ \tcoal$, the size of the entire projected tier, which was therefore reading back the reference rather than the map. Every number below is instead scored against a solve at rtol $=10^{-14}$, whose residual error, measured as its difference from rtol $=10^{-13}$, is $1.0\times10^{-7}$~rad; we quote no arm below $10^{-7}$~rad, and the two arms that reach it are reported as truth-limited. Per-orbit arms are scored as $2\pi n + {\rm unwrap}$, with per-window excess over $2\pi$ below $7\times10^{-3}$~rad throughout. The classical reference arms are computed independently with each architecture and agree to four significant figures.

\subsubsection{Results}\label{sec:rr_results}
\begin{figure}[t]
\centering
\includegraphics[width=\columnwidth]{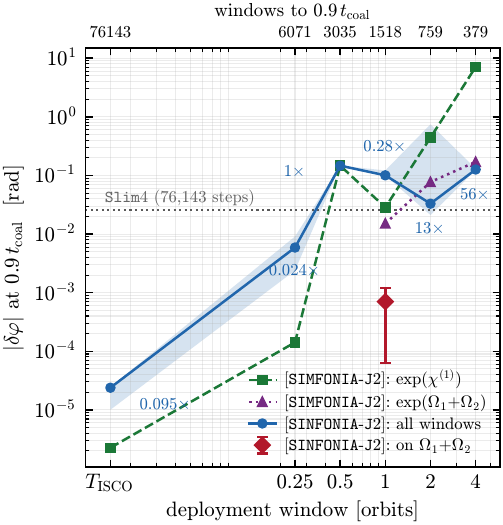}
\caption{Orbital-phase error $\Delta\phi$ at $0.9,t_{\rm coal}$ as a function of the deployment window advanced by each map evaluation. A single residual generator is trained over the range $t_{\rm ISCO}$--$4,t_{\rm orb}$ and then evaluated at six window sizes; points show the median over three seeds and the shaded band indicates the corresponding spread. The network-off control, $\exp(\chi^{(1)})$, is evaluated at the same windows. Each point represents a complete inspiral evolved to $0.9,t_{\rm coal}$, while the upper axis gives the corresponding number of map applications. Numerical labels indicate the ratio of the network-off to network-on phase error.}
\label{fig:ladder}
\end{figure}

We first assess the secular channel identified in Section~\ref{sec:arch_jx_secular_channel}. Across nine \texttt{J2} generators distilled against the exact one-window map, the pointwise generator error is \emph{anti}-correlated with the chained outcome: the best checkpoint, at $0.012$~rad chained error, has the largest pointwise error of the nine, while the worst, at $4.3$~rad, has a smaller one. What orders the models instead is the radially integrated signed energy-rate defect. Imposing this channel structurally through the shell projection, one force-law scalar per window, applied to already-trained weights, takes the same three checkpoints from $0.81$~rad with a $366\times$ band to $8.8\times10^{-5}$~rad with a $7\times$ band, without retraining (Table~\ref{tab:main}, Figure~\ref{fig:tisco}). The same intervention on a network-free order-7 Taylor tower reduces the chained phase error from $6.6\times10^{-3}$~rad to below $10^{-7}$, a gain of more than five orders of magnitude. At that level the measurement is limited by the reference solution rather than by the map. Applied to the analytic first-order Magnusian it similarly gives $4.1\times10^{-2}\ {\rm rad} \;\longrightarrow\;1.7\times10^{-7}\ {\rm rad}$ at $15\ \mu$s per window. Against \texttt{Slim4}'s $2.6\times10^{-2}$~rad, \emph{the projected analytic Magnusian is therefore more than five orders of magnitude more accurate while requiring only a few per cent of its evaluation cost.}

The mechanism is independent of the particular generator. Two controls establish what the projection is, and is not, doing. With $V=0$, the projection alone gives $2.2\times10^{3}$~rad and loses the orbit after $44\%$ of the inspiral: \emph{enforcing the secular budget without an accurate local propagator is not sufficient}. In contrast, when the same scalar \emph{is imposed only as a training penalty rather than as a structural projection}, the network satisfies the penalty while the chained map fails at $359$~rad. \emph{Arm \texttt{J1} reproduces the same finding with an independent architecture}: its learned remainder lands at $10^{-3}$--$10^{-1}$~rad with the projection disabled, and at $3.1\times10^{-6}$~rad when it is imposed (median of three seeds, band $4.4\times$, unsupervised). \emph{These interventions therefore isolate the signed secular projection, rather than pointwise map error alone, as the quantity controlling long-time rollout in this problem.}

We next assess the operator-side representation and what its dictionary buys at the present level of approximation. Fitting equation \eqref{eq:j3} by singular-value least squares on $32{,}768$ states in the training annulus and sweeping the dictionary degree separates the pointwise and secular error channels particularly cleanly. Without projection, the chained phase error falls by almost five orders of magnitude as the dictionary is enlarged, $\{5.4,\quad 1.04,\quad 0.17,\quad 1.07\times10^{-2},\quad 9.8\times10^{-5}\ {\rm rad}\}$
for $d=\{2,4,6,8,10\}$, corresponding to $10$--$286$ features. The one-window relative fit residual falls simultaneously from $1.3\times10^{-2}$ to $3.4\times10^{-6}$. With the secular projection imposed, however, the same sequence becomes $ \{3.4\times10^{-5},\quad 2.4\times10^{-6},\quad 1.1\times10^{-6},\quad 6.1\times10^{-7},\quad 7.2\times10^{-7}\}\ {\rm rad}.$ Thus almost five orders of raw improvement are compressed into approximately one order once the secular channel is fixed, and the $d=10$ map is marginally worse than the $d=8$ map. Beyond the lowest useful dictionary orders, increasingly accurate pointwise representation therefore provides only marginal additional long-time accuracy once the dominant secular defect has been removed. This is the conclusion of Section~\ref{sec:arch_jx_secular_channel} reproduced on the operator side \emph{in a representation containing no gradient optimizer.} At $d=8$ the projected least-squares map reaches $6.1\times10^{-7}$~rad over three independent data draws, with a band of $1.0\times$ [$5.9$--$6.2\times10^{-7}$], at $6.8\,\mu$s per window. It is the cheapest arm in this work that lies within one order of magnitude of the reference floor. The corresponding controls behave as for the \texttt{J2} architecture: the identity interaction-picture map gives $2{,}504$~rad, while the projection with no dictionary gives $2{,}232$~rad and tracks only $43\%$ of the inspiral. Replacing the closed-form least-squares fit by Adam on the same dictionary reaches $1.4\times10^{-5}$~rad projected, while an MLP encoder reaches $8.1\times10^{-6}$~rad (three seeds each). Their unprojected errors, $0.77$ and $0.18$~rad respectively against $1.07\times10^{-2}$~rad for the least-squares map, \emph{indicate that the gap is primarily one of optimization rather than representation}. At this window, neural optimization therefore does not improve the closed-form operator fit, \emph{what it can buy is evaluation cost}, $5.2 \mu$s instead of $6.8\ \mu$s.

A separate result concerns the Magnusian itself. Chaining the single-bracket update $ z\mapsto\Phi^K\!\big(z+\chione(z)\big),$ our own starting point and not the construction of \citet[Section 4.3]{blanco2026magnusian}, who exponentiates the generator and retains nested brackets, gives $4.1\times10^{-2}$~rad at $\tisco$ independently of quadrature refinement. Chaining the same generator as a genuine flow, $ z\mapsto\Phi^K\!\big(\exp(\chione)(z)\big),$
reduces the error to $2.3\times10^{-6}$~rad: approximately four orders of magnitude below \texttt{Slim4}, below adaptive \texttt{RK45} even after $3\times10^{6}$ force evaluations, \emph{and with nothing learned}. \emph{At one orbital period per step, the same exponentiated analytic map gives $2.85\times10^{-2}$~rad in only $1{,}518$ windows, essentially \texttt{Slim4}'s phase accuracy at fifty times fewer map applications and a wall time of $0.8$~s.}

Figure~\ref{fig:perorbit} then tests the learned generator in this more demanding one-orbit regime. At $r_0=100\ M$, the plain residual network ends $1.3\times$ below the exponentiated first-order closed form (median of five seeds, band $2.7\times$), and $3$--$30\times$ below it through the first half of the inspiral. The secular-aware fine-tune improves the median to $2.8\times$ below the closed form, or $1.0\times10^{-2}$~rad. \emph{This is below \texttt{Slim4}'s $2.6\times10^{-2}$~rad while advancing the orbit fifty times further per map application.} The best seed reaches $1.6\times10^{-3}$~rad, $18\times$ below the analytic first-order map, and three of five seeds improve. The secular-aware term is the first training-time objective used here whose diagnostic tracks the chained error: its predicted secular defect falls from $10^{-1}$--$10^{-2}$ to $10^{-3}$~rad, and the long-time phase error falls with it. The pointwise loss does not provide the same ordering.

Transfer at one orbit per step is partial. Deployed at $r_0=200\,M$ without retraining, the plain networks remain $1.5\times$ below the first-order closed form in the median of five seeds, with three seeds improving by factors of $3$--$4$ and two adverse. The secular-tuned networks are instead $3$--$6\times$ worse than the analytic map: the secular correction was fitted on the $100\ M$ panel and does not transfer unchanged. Training the correction on a panel spanning $\rr$ distributes the error more uniformly across separations rather than minimizing it at a single point: for one representative seed the errors are $2.4$, $2.1$, and $5.0\times10^{-2}$~rad at $r_0=100$, $200$, and $50\ M$, against $2.85$, $0.50$, and $16\times10^{-2}$~rad for the corresponding closed forms. At $r_0=50\ M$, every one-orbit arm reaches $0.12$--$0.17$~rad. Here the per-orbit $\chitwo$ content is approximately five times larger than at $100\ M$. \emph{The limitation is therefore the size of the finite window rather than an observed instability of the learned map. No arm becomes dynamically unstable over the tested range.}

The local-error scaling gives a complementary diagnostic. The shell-projected \texttt{J2} map at $\tisco$ has exponent $\simeq1.0$, and the plain per-orbit residual network $\simeq1.2$, comparable to \texttt{Slim4}'s $\simeq1.3$ under the same diagnostic. By contrast, fixed-step \texttt{RK4} and adaptive \texttt{RK45} give slopes $\simeq2.2$--$2.4$. The exponentiated analytic maps and the secular-tuned networks carry a small fixed-sign residual with an approximately quadratic scaling, but at substantially smaller amplitude. The scaling exponent and absolute error amplitude are therefore independent diagnostics, and both are reported in Table~\ref{tab:main}.

Figure~\ref{fig:ladder} isolates the role of the window itself by following a single residual architecture across window lengths. The window is the natural axis along which higher Magnus content becomes important: the relative size of $\chitwo$ increases with $\Delta t$, while $\chione$ remains analytically available. A single residual generator trained with the window sampled log-uniformly from $\tisco$ to four orbital periods fits the Magnus remainder to $0.8\%$ of its prior throughout the interval (three seeds, validation loss $2.3$--$2.6\times10^{-4}$), and the magnitude of its per-window correction tracks that of the true remainder from $\tisco$ to four orbits. When chained using the pointwise objective alone, however, the same weights improve on the exponentiated first-order closed form only once the window exceeds approximately one orbit. At four orbits per window, where $\exp(\chione)$ has degraded to $7.1$~rad, the learned map carries the full inspiral in only $379$ windows to $0.126$~rad (median of three seeds, band $1.06\times$): $56\times$ below its own network-off map, although still approximately five times above the $76{,}143$-step \texttt{Slim4} result. At two orbits the median gain is $13\times$, with one adverse seed. At one orbit and below, the same learned correction instead degrades the map by factors of $3.5$--$40$, and at half an orbit its effect is negligible.

The pointwise fit is therefore correct in magnitude throughout the ladder, while the sign of its secular projection is not controlled. The window sweep measures that distinction within a single trained model and reproduces the mechanism of Section~\ref{sec:arch_jx_secular_channel} without changing architecture or optimizer. The corresponding network-free curve is itself non-monotonic: $\{2.3\times10^{-6},\ 1.4\times10^{-4},\ 0.15,\ 0.028,\ 0.44,\ 7.1\}$~rad at $T_{\rm ISCO}$, $\tfrac14$, $\tfrac12$, $1$, $2$, and $4$ orbital periods respectively. Diagnostics in the co-rotating frame show that fractional-orbit windows retain a signed oscillatory contribution that can accumulate coherently, whereas integer-orbit windows permit substantially greater cancellation around the cycle. \emph{The non-monotonicity is therefore a property of chained secular accumulation at large finite windows rather than of the formal local order of the Magnus expansion.} It is invisible from a one-window derivation and marks the regime in which a learned higher-order generator can become useful. A secular-aware objective defined across window lengths is the natural extension.

Above one orbit, however, the fair analytic control is not $\exp(\chione)$ but the second-order Magnusian of Section~\ref{sec:arch_j2}. With no learned residual, $\exp(\Omega_1+\Omega_2)$ gives $1.5\times10^{-2}$, $7.7\times10^{-2}$, and $0.17$~rad at one, two, and four orbits respectively: factors of $1.9$, $5.7$, and $41$ below $\exp(\chione)$. To our knowledge, this is the first long-duration chained test of the second-order dissipative Magnusian on an inspiral, we will give more details on the completion of this piece in \citet{Magnusian2}. Against this stronger analytic control, residual networks trained on the first-order base sit only $1.3$--$2.3\times$ lower, at $1.0\times10^{-2}$, $3.3\times10^{-2}$, and $0.126$~rad at one, two, and four orbits. Except at the largest window these differences lie within the corresponding seed bands. Thus, on the first-order base, the learned residual predominantly supplies content that the explicit second-order Magnusian already captures. The decisive test is therefore to place the residual network on the second-order base, so that its learned correction can act only on the content remaining beyond $\Omega_2$. With the residual prior rescaled to this smaller remainder and otherwise the same training procedure, three seeds of the plain distilled model chain to $2.8\times10^{-3}$~rad (band $1.25\times$), a factor of $5.4$ below $\exp(\Omega_1+\Omega_2)$, with every saved checkpoint outperforming the network-off second-order map. The secular-aware fine-tune reaches a median of $7.0\times10^{-4}$~rad (three seeds, band $19\times$), with the best seed at $6.4\times10^{-5}$~rad: $21\times$ below the second-order analytic Magnusian and $37\times$ below \texttt{Slim4} while using one fiftieth of its window count. \textbf{This is the network contribution measured against the strongest analytic map constructed here: what remains beyond the explicit second-order generator is learned by the residual.}

At the much shorter $\tisco$ window, by contrast, \emph{out-of-domain transfer is substantially stronger.} The \texttt{J1} and \texttt{J2} networks trained at $r_0=100\ M$ and deployed without retraining at $200\  M$ and $50\ M$ against fresh \texttt{DOP853} references ($1.2\times10^{6}$ and $4{,}759$ windows, respectively) track the entire evolution. At $200\,M$, all analytic and learned projected arms reach a common $4.7\times10^{-5}$-rad floor. At $50\ M$, where the corresponding $\tisco'$ lies $2.8\times$ outside the trained range, the projected learned maps remain at the level of their projected analytic counterparts ($2.5\times10^{-5}$ versus $2.2\times10^{-5}$~rad). Thus the transfer seen at the short window is carried primarily by the analytic structure, which is recomputed at the new $\rr$; the learned residual remains bounded and, in these tests, does not degrade it. \emph{This is an empirical result over the range tested here, not a general guarantee.}

\subsection{A hidden environment: dynamical friction}\label{sec:hidden_df} 
We now ask whether a force \emph{hidden} from the model, present in the dynamics that generated the data but absent from the propagator used to fit them, can be recovered from the chained phase alone. This inverts Section~\ref{sec:rr_results}: there the secular channel was imposed from a known flux; here it is left free, and the question is whether it is identifiable. We choose a dynamical-friction-like drag as the test force for two reasons. First, dynamical friction is a canonical environmental effect for LISA sources, arising for example in dark-matter spikes and gaseous environments. Second, for the quasi-circular motion considered here it is velocity-aligned and therefore lies within the single secular channel available to the freed scalar. It thus provides a sharp in-class test of the channel as an inference variable, while the radial force introduced below serves as an out-of-class control.

\subsubsection{The force}\label{sec:drag_force} 

A body of mass $m$ moving through a medium of density $\rho$ feels a dynamical-friction drag \citep{chandrasekhar1943dynamical, ostriker1999dynamical}. For a dark-matter spike or an accretion disc with $\rho = \rho_0\, r^{-\gamma}$, on a quasi-circular orbit where $v^{2} = 1/r$, the supersonic limit of that force reduces to a phenomenological power law in the separation,
\begin{equation}
\mathbf{F}_{\rm df} = -\,\kappa\, r^{-p}\,\mathbf{v}, \qquad p = \gamma - \tfrac32,
\qquad \kappa = 4\pi G^{2} m^{2}\rho_0 \ln\Lambda ,
\label{eq:drag_power}
\end{equation}
up to the order-unity Coulomb factor, and whose dephasing of an inspiral is a target for LISA
\citep{barausse2014can, kavanagh2020detecting, speri2023probing}. Equation \eqref{eq:drag_power} is used here as a representative member of the velocity-aligned class rather than a calibrated astrophysical model: the Mach-number structure of the underlying Chandrasekhar--Ostriker force, the slow logarithmic growth of $\ln\Lambda$, and the mass-ratio scaling of an intermediate-mass-ratio system are all set aside, and the injection below is synthetic and noiseless. Our goal is to assess identifiability, not astrophysical fidelity.

Its energy-loss rate on a circular orbit of radius $a$, the quantity the secular channel of
Section~\ref{sec:arch_jx_secular_channel} actually carries, is
\begin{equation}
\dot E_{\rm df} = \mathbf{F}_{\rm df}\!\cdot\!\mathbf{v} = -\,\kappa\, a^{-(p+1)} .
\label{eq:drag_rate}
\end{equation}
Two properties of equation \eqref{eq:drag_power} decide what follows. First, it is velocity-aligned, $\mathbf{F} \parallel \mathbf{v}$, and any such force satisfies the circular-orbit flux lock of equation \eqref{eq:fluxlock} \emph{identically}, not merely to leading order
($dL_z/dE = r\,F_\theta/(\mathbf{F}\!\cdot\!\mathbf{v}) = r/v = 1/\omega$ exactly on a circle): it therefore lies inside the one-channel class of Section~\ref{sec:arch_jx_secular_channel} and is representable by a single secular scalar per window. Second, a force outside that class, a radial term, for instance the enclosed mass of the same medium, does not move $(E,L_z)$ on a circular orbit at all, and acts on the phase only through $\omega(r)$. The two cases are separated below, and they behave differently. We inject equation \eqref{eq:drag_power} into the truth with $p=1$ ($\gamma=5/2$) and $\kappa = 3\times10^{-8}$, worth $9.3$~rad of dephasing at $0.9\ \tcoal$, and hide it from the model.

\subsubsection{Results}\label{sec:drag_force_results}

\emph{Imposed, the secular scalar makes the map an integrator. Left free, it makes the map an instrument.} The shell projection takes the inspiral rate from a known flux and can therefore discover nothing. The same scalar, fitted to phase data with the remaining dynamics held fixed, can. We perform that fit on the network-free \texttt{[SIMFONIA-J2]} map, over the full $1{,}518$-window phase history ($1{,}517$ windows complete before $0.9\ \tcoal$ and are scored). The vacuum propagator contains no environmental force, and the missing rate is assumed only to lie in the two-parameter family $\dot E_\theta(a) = -\kappa a^{-(p+1)}$, applied as an exact $(E,L_z)$ shift along the circular flux lock given by equation \eqref{eq:fluxlock}. The injected drag of Section~\ref{sec:drag_force} ($\kappa=3\times10^{-8}$, $p=1$) is worth $9.3$~rad by $0.9\ \tcoal$. The model is told nothing about it.

\begin{figure*}[t]
\centering
\includegraphics[width=0.96\textwidth]{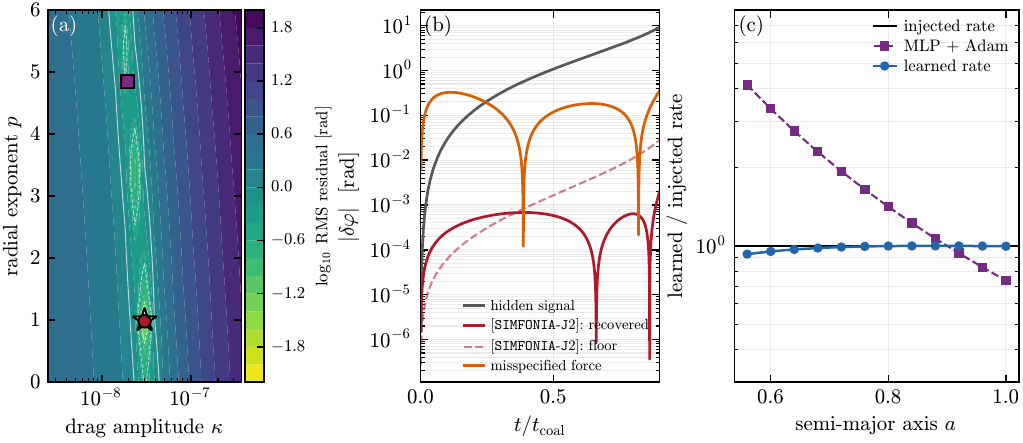}
\caption{Recovery of a hidden dissipative force from the accumulated orbital phase using the network-free \texttt{SIMFONIA-J2} map, with one application of $\exp(\chi^{(1)})$ per orbit and no secular projection. Only the secular rate is inferred: parametrically in panels (a)--(b) and as a free function in panel (c). (a) Root-mean-square phase residual of the two-parameter model over $(\kappa,p)$ relative to the injected drag evolution. The injected parameters (star) coincide with the global minimum; gradient descent initialized far from the solution converges to the degeneracy ridge (square), whereas a direct root solve recovers the injected point (circle). (b) Accumulated dephasing of the vacuum map relative to the drag reference, showing the $9.3$-rad hidden signal; of the recovered map relative to the same reference; and of the vacuum map relative to the vacuum reference, which sets the numerical integration floor. (c) Recovered secular rate when no functional form is supplied, shown as the ratio to the injected rate over the separations traversed by the inspiral. An MLP trained on a steep prior with a first-order optimizer converges to the ridge, whereas a
log-rate parametrization optimized with a quasi-Newton method recovers the injected rate.}
\label{fig:recovery}
\end{figure*}

Figure~\ref{fig:recovery} shows the residual landscape in $(\kappa,p)$. It contains a narrow valley whose minimum lies close to the injected parameters, and a derivative-free solver converges to the same basin from three separate starts. The recovered amplitude agrees with the injection to ${\sim}0.1\%$ and the radial exponent differs by $-1.8\%$, reducing the $9.3$-rad dephasing to $2\times10^{-3}$~rad. That residual should not itself be read as the recovery accuracy, since the fitted secular correction can absorb part of the deterministic propagator bias. The parameter recovery and the vacuum null are the relevant diagnostics. Fitting the same model to the vacuum truth returns an amplitude consistent with zero at the resolution of the optimization, $\kappa/\kappa_{\rm inj}\le2\times10^{-10}$. The power law itself need not be prescribed: replacing it by a small neural representation of $\log[-\dot E_\theta(a)]$ as a free function of $\log a$, retaining only the assumption that the missing physics lies in the same secular channel, recovers the injected rate to ${\sim}0.3\%$ in amplitude and leaves $1.1\times10^{-4}$~rad (Figure~\ref{fig:recovery}c).

The two parameters are limited by different effects. The amplitude is sensitive to forward-model systematics, and in particular to a cross term the model omits: $\chione$ is linear in the force, whereas the map is its exponential, so $\exp(\chi_{\rm vac}+\chi_{\rm drag}) \neq \Pi[\exp(\chi_{\rm vac});\Delta E,\Delta L_z]$, and representing the drag by a separate secular shift cannot carry the difference. Measured against exact flows, that difference is $7\times10^{-4}$ to $3\times10^{-3}$ of the drag's own one-window $\Delta E$ across the inspiral, and moving the trial drag inside the generator shifts the recovered $\kappa$ by $0.10\%$. We therefore regard the amplitude as accurate at the ${\sim}0.1\%$ level and attach no significance to smaller differences.

The exponent is limited by a different mechanism. The landscape contains an extended ridge in
$(\kappa,p)$: over a finite range in separation, a change in amplitude is partly compensated by a change in radial slope. A forward model whose minimum residual is $1.5\times$ smaller consequently returns a slightly worse $p$ ($-3.0\%$ rather than $-1.8\%$), the minimum having moved along the ridge rather than the ridge having narrowed. What breaks the degeneracy is the inspiral's own sweep in separation, which is why the full chained history is informative and why $p$ is the more fragile of the two.

Finally we test the model outside the class it is built to represent, adding to the injected drag a radial perturbation $\mathbf{F}_r = -\beta r^{-3/2}\hat{\mathbf r}$ with $\beta=5\times10^{-4}$, the enclosed mass of a $\rho \propto r^{-5/2}$ medium, and fitting the pair with the same velocity-aligned secular model. The scalar correction absorbs much of the resulting phase shift, reducing a $36$-rad dephasing to $0.35$~rad, but does not reproduce the perturbation consistently: from three separate starts the fit converges to a drag amplitude $27$ times the injected one and to $p=-51$. Two diagnostics expose this without knowledge of the injected parameters. First, the recovered parameters are unphysical: $p=-51$ implies a dissipation growing extremely rapidly with
separation, incompatible with the drag model being fitted. Second, the residual sits an order of magnitude above the map's own $2.85\times10^{-2}$-rad floor (the vacuum map's dephasing against the vacuum truth, the per-orbit $\exp(\chione)$ accuracy of Table~\ref{tab:main}), where the well-specified recovery sits an order below it, a factor of $170$ between the two residuals (Figure~\ref{fig:recovery}b). A small residual does not by itself validate the inferred physics, it must be read together with the forward-model floor and the recovered parameters.

The conclusion is deliberately limited. All vacuum parameters are held fixed and known here. The degeneracies between environmental and intrinsic parameters that dominate realistic inference are outside this experiment, as are noise and a likelihood. In this controlled, noiseless setting, with the known vacuum dynamics held inside the analytic generator, missing dissipative physics is recoverable from the chained phase history when it is confined to the single secular channel identified above, either within a prescribed parametric family or as a free function of separation over the interval swept by the inspiral. Velocity-aligned effects such as dynamical friction and gas drag on quasi-circular orbits belong to this class. The radial test shows that physics outside it can nevertheless be partially absorbed by an effective secular rate, but only at the cost of pathological inferred parameters and a residual well above the calibrated forward-model floor. The
question posed at the start of this section is thus answered in this controlled setting: a force hidden from the model \emph{is} recoverable from the chained phase alone, to $0.1\%$ in
amplitude and $2\%$ in exponent when it lies in the secular class the instrument represents, with a clean null on the vacuum, and with two audit diagnostics, unphysical parameters and a residual far above the calibrated floor, that expose a force outside that class without knowledge of the injection.

\section{Discussion} \label{sec:discussion}

We have developed a family of classical \texttt{[SIMFONIA]} and neural \texttt{[SINFONIA]} finite-flow maps for conservative and dissipative orbital dynamics that combine the interaction picture of the exact Kepler-Flow with Taylor/Magnusian generator, impose the map's singular secular channel structurally from the force law, and admit a learned residual generator on top.  

Three learned flow maps for a dissipative system, all trained from the force law alone, were chained through a complete radiation-reaction inspiral against the integrator that defines the benchmark, with the network-off control in every row. We found \emph{adding a secular channel crucial}: the chained error of any fixed-window map on this system is one signed scalar per window, imposing it structurally takes analytic and learned generators alike to the $10^{-6}$-rad tier, and the pointwise objectives every architecture was trained with are provably blind to it. Second, \emph{the exponentiated first-order Magnusian} is itself a $10^{-6}$-rad integrator at that step and matches \texttt{Slim4} one orbit per step, with no learned part. Third, \emph{a learned Magnusian generator} on that structure carries the inspiral one orbit per step, improves on the closed form where it was trained ($2.8\times$, median of five), transfers across separation with the analytic map in the family and the deviation bounded, and \emph{is slimplectic} in Tsang's \citet{galley2013classical, tsang2015slimplectic} sense. Fourth, \emph{the channel cuts both ways: pinned, it sets the floor; left free, it is the one quantity a data-driven map must learn}, and a dynamical-friction-like drag hidden from the model is recovered from the chained phase alone, amplitude and radial exponent, by a two-parameter rate and by a learned function of the separation, with the vacuum fit returning zero.

The gain in using a neural flow therefore does not come from replacing an analytic map that is already sufficiently accurate. At the benchmark step the exponentiated first-order Magnusian leaves too little unresolved structure for the learned residual to improve upon reliably. \emph{Learning becomes useful only as the finite window grows and higher-order generator content becomes dynamically resolvable}. The window ladder makes this transition explicit, while the second-order control identifies its origin: on a first-order base the residual network largely supplies content already contained in $\Omega_2$, whereas on the explicit second-order base the remaining residual is again learned and improves the strongest analytic map available. This suggests \emph{a practical architectural design rule for the programme}: resolve analytically, or by structure-preserving construction, every component that is cheaper and more accurate to prescribe, and amortize only the finite-time content that remains.

The conclusion is nevertheless specific to the regime tested here. The benchmark is quasi-circular, planar, non-spinning, and deliberately weak-field. Its physical dissipative drift collapses to a single secular channel, and no transient multi-frequency resonance is encountered. Neither the persistence of this one-dimensional reduction nor the observed transfer properties should therefore be assumed for generic eccentric, precessing, resonant, or strong-field motion. Likewise, the shell projection is effective here because the relevant balance structure is known from the force law. \emph{In problems with several independent secular directions, the corresponding constrained subspace must first be identified.} The present results should therefore be read as establishing the mechanism and the architecture under controlled conditions, rather than as a production waveform model. 

The present calculations use deliberately simple orbital models \emph{in order to isolate} the numerical and architectural mechanisms controlling long-time rollout, but the construction is not tied to this level of approximation. Higher-order post-Newtonian dynamics provide a direct next step, including spin couplings and higher-order radiation reaction, while retaining the same structure-preserving and equation-based training strategy. A particularly natural target is effective-one-body (EOB) dynamics, where accurate conservative and dissipative equations of motion are available but closed-form finite-time solution maps are generally not. In this setting, a learned structure-preserving flow amortizes the repeated integration of an increasingly expensive EOB vector field while retaining the known Hamiltonian and dissipative structure.

This perspective is reinforced by a central feature of \texttt{[SINFONIA-JX]}, as highlighted in Section \ref{sec:arch_jx_secular_channel}: the framework is \textbf{trajectory-unsupervised}. \emph{No reference inspirals are used during training.} The learned maps acquire the conservative geometry and dominant dissipative evolution directly from the governing equations. Consequently, future supervised information need not be used to reconstruct the full dynamics from scratch. \emph{Instead, sparse or expensive reference data could be concentrated on physics that is absent from the analytic model, uncertain, or prohibitively costly to incorporate directly—for example higher-order self-force effects, numerical-relativity corrections, environmental perturbations, or waveform-level residuals.} This suggests a natural hierarchy in which PN or EOB theory supplies the dynamical backbone, while supervision is reserved for the genuinely unresolved part of the problem. Updates to our current \textbf{classical and neural} \texttt{[SIMFONIA||SINFONIA]} research programme will continually be provided \citet{GomesdaSilvasinfoniaWP} along with is respective repositories. 

\begin{acknowledgments} 
We would like to thank Rohit Chandramouli for reviewing this draft, for key discussions, and for collaboration on upcoming works in this programme. We further thank the participants of the GWSky kick-off \& Sexten meeting for discussions and
collaboration, and, of course, the Magnusian squad at the AEI. We are especially grateful to Francisco Blanco: his Magnusian construction underpins the third architecture of this work, and his insight and correspondence have been crucial throughout. We would also like to thank Davide Murari for extensive discussions on both geometric numerical time integration and machine learning, and Sid Mahesh for discussions on BOB and alternative neural approaches to the improvement of EOB--BOB models. Finally, we thank Christian Lubich and Bernd Br\"ugmann for clarifications on their beautiful work \citep{lubich2010symplectic, heinze2026n}, which has been a great source of inspiration for our research programme. Anthropic's \texttt{Claude} was used for assistance with code development and numerical optimization, and manuscript editing; the author is solely responsible for all content. The author is funded by the European Union (ERC grant GWSky/ 101167314) and the PRIN 2022 grant \enquote{GUVIRP - Gravity tests in the UltraViolet and InfraRed with Pulsar timing}. Views and opinions expressed are however those of the author(s) only and do not necessarily reflect those of the European Union or the European Research Council Executive Agency. Neither the European Union nor the granting authority can be held responsible for them. \end{acknowledgments}

\software{\texttt{JAX, EQUINOX, OPTAX, SymPy, NumPy, SciPy, Matplotlib.}}\\

\bibliography{sinfonia-p0}{}
\bibliographystyle{aasjournalv7}

\end{document}